\documentclass[sigconf]{acmart}
\acmConference[xxxx 2024]{XXXX Conference}{1 May, 2024}{Shanghai, China}

\usepackage{algorithm}
\usepackage{algorithmic}
\usepackage{multirow}
\usepackage{adjustbox}
\usepackage{tabularx}
 \usepackage{booktabs}
 \usepackage{graphicx}
\usepackage{array}

\AtBeginDocument{%
  \providecommand\BibTeX{{%
    \normalfont B\kern-0.5em{\scshape i\kern-0.25em b}\kern-0.8em\TeX}}}

\setcopyright{acmlicensed}
\copyrightyear{2024}
\acmYear{2024}
\acmDOI{XXXXXXX.XXXXXXX}

\acmBooktitle{Conference Paper} 
\acmISBN{978-1-4503-XXXX-X/18/06}

\begin{document}

\title{Risk Scenario Generation for Autonomous Driving Systems based on Causal Bayesian Networks}


\author{Jiangnan Zhao}
\affiliation{%
\institution{Software Engineering Institute}
  \institution{East China Normal University}
  \city{Shanghai}
  \country{China}}

\author{Dehui Du}
\affiliation{%
\institution{Software Engineering Institute}
  \institution{East China Normal University}
  \city{Shanghai}
  \country{China}}

\author{Xing Yu}
\affiliation{%
\institution{Software Engineering Institute}
  \institution{East China Normal University}
  \city{Shanghai}
  \country{China}}

\author{Hang Li}
\affiliation{%
\institution{Software Engineering Institute}
  \institution{East China Normal University}
  \city{Shanghai}
  \country{China}}

\renewcommand{\shortauthors}{Jiangnan Zhao and Dehui Du, et al.}

\begin{abstract}
Advancements in Autonomous Driving Systems (ADS) have brought significant benefits, but also raised concerns regarding their safety. Virtual tests are common practices to ensure the safety of ADS because they are more efficient and safer compared to field operational tests. However, capturing the complex dynamics of real-world driving environments and effectively generating risk scenarios for testing is challenging. In this paper, we propose a novel paradigm shift towards utilizing Causal Bayesian Networks (CBN) for scenario generation in ADS. The CBN is built and validated using Maryland accident data, providing a deeper insight into the myriad factors influencing autonomous driving behaviors. Based on the constructed CBN, we propose an algorithm that significantly enhances the process of risk scenario generation, leading to more effective and safer ADS.
An end-to-end testing framework for ADS is established utilizing the CARLA simulator. Through experiments,  we successfully generated 89 high-risk scenarios from 5 seed scenarios, outperforming baseline methods in terms of time and iterations required.

\end{abstract}




\maketitle

\section{Introduction}
With advancements in core technologies such as perception, computation, and control, autonomous vehicles have made significant strides in recent years. Major automotive companies and tech firms are actively pushing forward the commercialization of autonomous vehicles. According to the SAE standards\cite{SAEJ3016}, autonomous driving technology has achieved preliminary Level 3 conditional automation. Full self-driving technology is also steadily progressing. Despite the notable achievements in autonomous driving technology, accidents caused by errors in Autonomous Driving Systems (ADS) continue to occur, and many studies have reported vulnerabilities in existing ADS\cite{fengDenseReinforcementLearning2023,huaiDoppelgangerTestGeneration2023,defectsADSgap}.

Ensuring the safety and reliability of ADS is of paramount importance. The robustness and fault tolerance of these systems in complex and dynamic real-world road environments directly impact their feasibility. Therefore, apart from algorithm development, testing and verification of ADS are crucial. Testing helps discover defects, risks, and faults in the system, contributing to its continuous improvement and optimization.
As one of the testing methods, field operational testing is expensive and cannot cover a wide range of extremely complex scenarios. Effectively generating realistic risk scenarios to test ADS is a significant technical challenge facing the industry\cite{tangSurveyAutomatedDriving2023a}. 

Currently, scenario generation for ADS faces multiple unresolved challenges. A primary issue is the reliance on specific rules or expert experience for constructing test scenarios \cite{ontologygeneration, dingSurveySafetyCriticalDriving2023}. This approach introduces subjectivity and limits the ability to encompass a wide range of complex scenarios. Additionally, the mechanisms employed in existing testing frameworks for scenario generation are simplistic and time-consuming, often relying on random mutation of scenario element parameters\cite{dingSurveySafetyCriticalDriving2023}, such as arbitrarily altering weather and lighting conditions. Furthermore, the assessment of driving behavior tends to be uni-dimensional\cite{safe_to_drive_metric}, lacking the depth needed to accurately reflect the nuances of real-world driving behavior.

In response to the various issues present in existing autonomous driving testing, we propose a novel framework for autonomous driving scenario generation based on \textit{causal Bayesian Networks (CBN)}, as shown in Fig.~\ref{framework}. The main innovative work includes:
\begin{itemize}
 \item Building a CBN of driving scenarios based on real accident data. Through data mining and causal inference algorithms, key risk factors associated with environmental conditions and behavioral patterns related to accidents are automatically identified, reducing subjective assumptions during the construction of test scenarios.
\item Defining CBN-based scenario generation strategies. Compared to baseline methods, our approach can effectively and efficiently generate risk scenarios in fewer iterations.
\item Introducing a comprehensive driving behavior assessment method. It evaluates driving behaviors from multiple dimensions, including safety, comfort, and efficiency, offering a more holistic understanding and evaluation of driving performance.
\end{itemize}

The rest of this paper is organized as follows. 
Section II establishes the background on scenario-based testing and CBN.
Section III describes a detailed related work of risk scenario generation for ADS.
Section IV elaborates on the overall design concept and key scenario-generation processes of the framework proposed in this paper.
Section V evaluates the effectiveness, efficiency, and runtime performance of the framework.
Section VI discusses internal and external threats.
Section VII concludes the paper.

\section{Background}

\subsection{Scenario-based testing for ADS}
Scenario-based testing of ADS aims at detecting unsafe behaviors of ADS in simulated driving scenarios. A scenario \cite{scenariosdefine} can be defined as the temporal development between several scenes in a sequence of scenes, where each scene snapshot captures the instantaneous state of all static and dynamic elements at that moment. According to ISO standards \cite{ISO3450}, autonomous driving scenarios can be categorized into functional scenarios (abstract), logical scenarios, and concrete scenarios (specific implementation).

Functional scenarios refer to the highest level of abstract scenario description that only focuses on the main functions and behaviors of the scenario without specific implementation details. For example, "a vehicle driving at a stable speed on a highway".

Logical scenarios enrich functional scenarios by adding parameters, and environmental effects to form intermediate descriptions required for training and testing. In the above example, elements like speed range, weather conditions, etc. can be specified, yet still expressed abstractly.

Concrete scenarios convert logical scenario descriptions into specific formats that can be understood by testing platforms. This requires the randomization of scenario parameters, insertion of traffic flows, events, etc. A scenario that can be directly loaded and executed in testing platforms.

Scenario-based testing can effectively discover defects, risks, and failures of ADS, which is crucial for ensuring their safety and reliability. Compared with conventional road tests, it offers more cost-effective ways to cover extensively complex scenarios. However, how to generate realistic risk scenarios remains a technical challenge.

\subsection{Causal Bayesian Networks}
A Causal Bayesian Network \cite{bookPearlMackenzie18} is a refined subset of Bayesian Networks \cite{bookPearlMackenzie18}, distinguished primarily by its explicit representation of causal relationships. This advanced probabilistic graphical model depicts variables and their conditional dependencies through a directed acyclic graph (DAG). However, unlike a standard Bayesian Network that models probabilistic relationships without direct causal implications, a CBN explicitly encodes causality in its structure.

The edges in a CBN are not merely indicative of associative relationships; they represent direct causal influences. This characterization is crucial in scenarios where understanding the impact of manipulating one variable on another is of paramount importance. Therefore, CBN finds extensive use in fields like epidemiology, economics, and policy analysis, where assessing the outcomes of interventions is critical.

In contrast to traditional Bayesian Networks, which excel in probabilistic inference and prediction, CBN facilitates causal inference, enabling researchers and practitioners to draw conclusions about cause-and-effect relationships. This distinction is vital in the context of experimental design, decision-making, and scenario analysis, where understanding the directionality and influence of relationships is as important as understanding the relationships themselves.  

In autonomous driving scenario generation, CBN can effectively model factors involving humans, vehicles, and the environment. They facilitate bidirectional predictive analysis and identify external, unknown risks, offering a more comprehensive approach than methods focusing solely on inherent system risks. Hence, their use in driving scenario modeling is highly suitable.
\section{Related work}
\textbf{Testing autonomous driving systems. }In pursuit of ensuring the safety of Autonomous Driving Systems (ADS), substantial efforts have been made in both academia and industry \cite{ISO26262, ISO21448, khatunScenariobasedExtendedHARA2020}, encompassing risk analysis \cite{chiaRiskAssessmentMethodologies2022, liDatadrivenBayesianNetwork2023} prior to system development and testing and validation post-development \cite{fengTestingScenarioLibrary2021}. In the testing domain, many studies have focused on single aspects such as sensing \cite{sensing_app9112341}, perception \cite{percep_deepbillbord, percp_deeproad, percp_deeptest}, or planning \cite{plan_avoid}. For example, introducing noise into perception layer images to test the robustness of network models. Diverging from these approaches, our work considers ADS as an integrated system, conducting end-to-end tests that closely align with real-world usage. This holistic testing approach is more adept at uncovering systemic flaws.

\textbf{Risk scenario generation. }Recently, attention has also shifted to whole-vehicle testing, as exemplified by Waymo \cite{waymoOpenDataset} and Baidu Apollo's \cite{baiduApollox5F00x53D1x8005x793Ex533A} trials on closed roads and in simulation environments. However, closed-road testing is limited by regulatory restrictions and rarely encounters risk scenarios. To address this, virtual testing \cite{dingSurveySafetyCriticalDriving2023,tangSurveyAutomatedDriving2023a} has emerged as a focal area. The core challenge in virtual testing lies in scenario generation \cite{nalicScenarioBasedTesting2020}, with existing methods categorized as search-based \cite{ontologygeneration, wangRiskScenarioGeneration2023}, adversarial \cite{fengDenseReinforcementLearning2023}, mutation-based \cite{liAVFUZZERFindingSafety2020, kimDriveFuzzDiscoveringAutonomous2022, gene_lawbreaker}, and accident data-based approaches \cite{gene-data-1}. Search-based methods primarily examine interactions between road objects, using optimization algorithms to pinpoint critical corner cases. Adversarial approaches alter the behaviors of background vehicles, forcing the ADS to handle dangers. Mutation-based methods employ genetic algorithms or specific rules to mutate scenario attributes. There's also growing research in reconstructing dangerous scenarios from accident data. However, these methods often require multiple iterations or are tailored to specific scenarios, failing to adequately represent the complexities and dynamics of the real world.

In contrast, our paper constructs a CBN from real accident data, enabling a more accurate capture of actual environmental features. Our method targets higher-level functional scenarios, expanding the scope of testing. With minimal iterations, it generates a multitude of risk scenarios, demonstrating superior efficiency in system evaluation.

\section{Methodology}
\begin{figure}
\centerline{\includegraphics[width=0.5\textwidth]{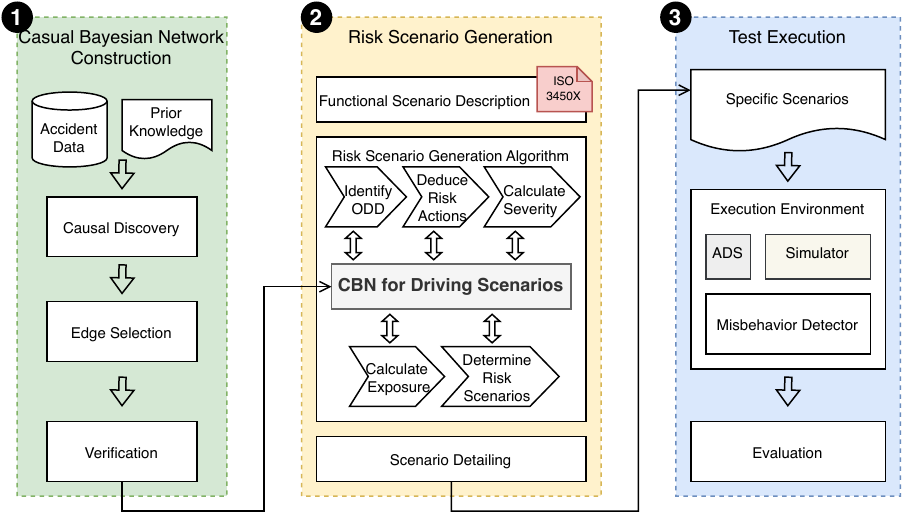}}
\caption{Overview of our approach}
\label{framework}
\end{figure}

In this section, we present a framework for autonomous driving scenario generation based on CBN. It aims to guide the generation of autonomous driving scenarios based on CBN and thus can generate driving scenarios that can lead to incidents more effectively. Fig.~\ref{framework} illustrates the workflow of the proposed approach, which consists of three main modules: CBN construction, scenario generation, and test execution.

\subsection{The construction of CBN}\label{CBN}
\begin{figure}[htbp]
\centerline{\includegraphics[width=0.5\textwidth]{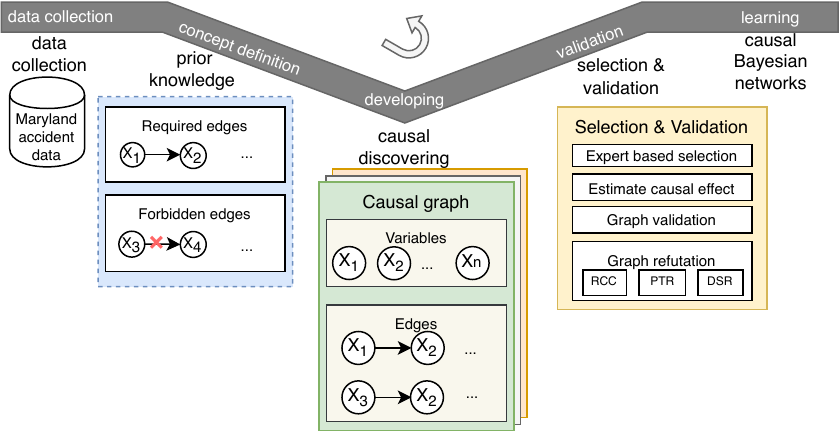}}
\caption{The pipeline of constructing CBN}
\label{pipeline}
\end{figure}

Inspired by the software engineering Verification and Validation (V\&V) model \cite{maValidationVerificationDomainspecific2023}, we have standardized the construction process of the CBN into five distinct stages, as depicted in Fig.~\ref{pipeline}. The stages are as follows:

(1). \textbf{Data collection phase:} Accident data consists of real-world records and simulated cases. While simulation offers comprehensive coverage of all driving-related metrics, it fails to capture real-life complexity and dynamics \cite{data_source_GUTIERREZOSORIO2020432}. In contrast, real-world data contains abundant authentic environmental details, though varying in quality. To achieve more realistic scene characterization, we opt for real crash records. After investigating multiple public datasets across countries \cite{data_catalog, calderdaleHomeCalderdale, marylandBenchmarkReports}, we selected the Maryland accident dataset \cite{marylandBenchmarkReports}. This dataset has ample scale with over one million incident cases statewide. It encompasses critical specifics like road states, weather, lighting, etc., forming a solid foundation for accurate driving scenario modeling. We further preprocessed the Maryland records to extract over 300,000 efficient accident data entries. Utilizing this data, we efficiently generate risk scenarios, reflecting the practical application of our approach in creating realistic and representative tests for ADS. 

(2). \textbf{Concept definition phase: }The algorithm is further informed by prior knowledge of causal relationships between specific variables. This knowledge, grounded in expert insights or logical deductions, is crucial in driving scenarios. For example, it's recognized that temporal changes influence light intensity variations. This addition of expert knowledge aids the algorithm in efficiently navigating through the potential causal structures, focusing on the most promising ones. Besides, the incorporation of specified and forbidden edges in the model narrows down the search area. This tailored approach not only accelerates the search process but also enhances the accuracy of the algorithm, enabling it to construct the most effective causal structure\cite{causal_discover_nogueira_methods_2022}.

(3). \textbf{Developing phase: }
During the causal discovery process in our study, we implemented specific parameter settings for each algorithm to optimize performance. This includes configuring algorithm parameters, such as the recursive depth in the Greedy Fast Causal Inference (GFCI) \cite{causal_discover, nogueiraMethodsToolsCausal2022} algorithm, to tailor it to the unique requirements of the causal model. An essential part of this process involves using testing and scoring methods to assess and select the most viable causal structures. Scoring techniques, such as the BDeu score \cite{bdeu}, are crucial for evaluating how well the model's parameters align with the data, considering variable dependencies, and incorporating smoothing techniques to avoid overfitting.
Additionally, we performed conditional independence testing, a critical step in refining the causal model. This involves applying various testing methods to the existing structures to determine whether to add or remove edges, which is vital for ensuring the model's accuracy and reliability. Moreover, common testing approaches such as G Square tests \cite{dowhy_gcm}, mutual information \cite{causal_discover}, and chi-squared tests \cite{causal_discover} were utilized to detect dependencies among variables in the data. These methods are indispensable in validating the model’s capability to accurately represent the underlying causal relationships.

(4). \textbf{Validation phase: }
In the validation phase of our causal graphs, we undertake the following crucial steps. Initially, based on expert knowledge, we select the most relevant edges and compute the causal effects between variables. To validate these causal graphs, we employ Structural Hamming Distance (SHD) \cite{dowhy_gcm} and Structural Interventional Distance (SID) \cite{dowhy_gcm} as our primary metrics. Our validation process also incorporates three distinct refutation methods to challenge the robustness of these graphs. The first method, Random Common Cause (RCC) \cite{dowhy_gcm}, introduces randomly selected covariates into the data and reanalyzes the changes in causal estimates, with minimal changes supporting our initial hypothesis. The second method, Placebo Treatment Refuter (PTR), assigns a random covariate as a treatment, expecting the new estimate to approximate zero if our hypothesis is valid. The third method, Data Subset Refuter (DSR), involves creating subsets of the data, akin to cross-validation, to assess the consistency in causal estimates across these subsets.

While these methods do not confirm the accuracy of the causal graph, they offer insights into its confidence level. As detailed in Table~\ref{tab:refutation}, the results indicate that the estimated effect and the new effect values are closely aligned, suggesting that the introduction of random confounding variables does not substantially alter the original causal effect. Furthermore, the p-values obtained from the Placebo Treatment Refuter (PTR) and Data Subset Refuter (DSR) methods are close to 0, reinforcing the high confidence level of our causal model. 

\begin{figure*}
\centerline{\includegraphics[width=\textwidth]{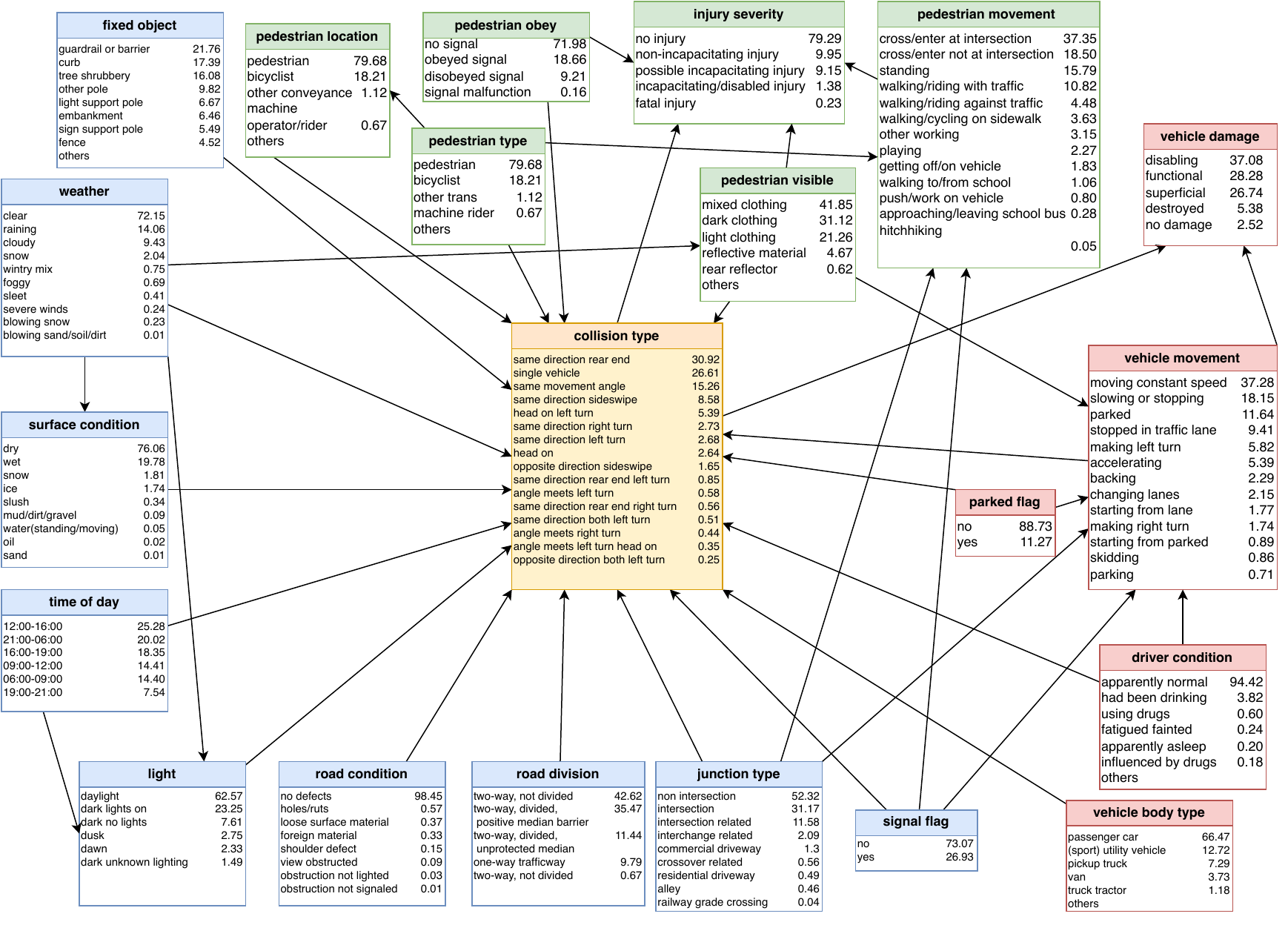}}
\caption{Causal Bayesian network for driving scenarios}
\label{model}
\end{figure*}

\begin{table}[htbp]
\caption{Refutation Results}
\label{tab:refutation}
\begin{tabularx}{\columnwidth}{@{}>{\hsize=5\hsize}X>{\hsize=13\hsize}X>{\hsize=3\hsize}X@{}}
\toprule
\textbf{Method} & \multicolumn{1}{l}{\textbf{Refutation Result}}                                                                   & \multicolumn{1}{l}{\textbf{Confidence}} \\ \midrule
RCC             & \begin{tabular}[c]{@{}l@{}}Estimated effect: 1853.532\\ New effect: 1801.223\end{tabular}               & High                           \\
\hline
PTR             & \begin{tabular}[c]{@{}l@{}}Estimated effect: 1632.98\\ New effect: -211.347\\p value: 0.421\end{tabular}  & High                           \\
\hline
DSR             & \begin{tabular}[c]{@{}l@{}}Estimated effect: 1655.654\\ New effect: 1632.211\\p value: 0.231\end{tabular} & High                           \\ \bottomrule
\end{tabularx}
\end{table}
(5). \textbf{Learning phase: }
After the validation of causal relationships in the autonomous driving scenarios, we engaged in an iterative process of parameter learning and structural learning based on accident data. This iterative approach allowed us to alternately refine the network's parameters and its structure, ensuring the alignment of the model with the underlying causal dynamics. Through this process, we were able to optimize and construct the best-fit CBN, as shown in Fig.~\ref{model}. The arrows represent causality, i.e., one variable influencing another. For example, “weather” is shown as a factor influencing “surface condition.” The numbers in the figure represent the probability of the various attributes.

\subsection{Scenario generation based on CBN}\label{generation}

After constructing the CBN, we can utilize it to generate driving scenarios. For a given functional scenario, delineated according to ISO driving scenario ontology \cite{ISO3450}, we perform a comprehensive deduction and analysis based on the CBN. Subsequently, risk scenarios are generated based on the results of the analysis.

Driving scenarios incorporate a variety of static elements. However, the environmental triggers of real-world traffic accidents are primarily concentrated in a few key areas: road obstacles, road surface lubricity, road damage, weather conditions, and lighting conditions. Based on this, we focus on these five critical factors to construct scenarios.
Additionally, unlike the random mutation of static elements common in previous methods, our approach employs cluster analysis to efficiently narrow down the search space and ensure the representativeness of the scenarios. Through cluster analysis, we have summarized the most frequently occurring static risk combination patterns in accident data.  Table \ref{tb:static} illustrates six of the discovered patterns. These combinations reflect the most common risk environmental compositions in the real world. For instance, a combination of rainy weather, slippery roads, and insufficient lighting is identified as a high-risk pattern. It should be noted that due to the limitations of the CARLA simulator in mimicking certain real-world scenarios, we implemented alternative measures. For instance, to simulate crosswinds, a significant risk factor identified in our accident data analysis, we applied lateral forces to all vehicles in the scenario, effectively replicating the conditions of strong wind weather. Using these patterns, we can effectively generate scenarios that cover typical risks. 
Finally, for each scenario element, we adopt a contract-based design philosophy, setting specific preconditions based on insights from the Maryland accident data and expert knowledge. This method helps in effectively reducing the number of iterations required for scenario generation, subsequently decreasing the frequency of scenario rendering in the simulator.

\begin{table*}[ht]
\caption{Static risk combination patterns}
\resizebox{\textwidth}{!}{%
\begin{small}
\begin{tabular}{@{}c|p{2cm}|p{2cm}|p{2cm}|p{2cm}|p{2cm}|p{2cm}p{2cm}@{}}
\toprule
\textbf{No.} & 1 & 2 & 3 & 4 & 5 & 6  \\ 
\midrule
\textbf{Category} & Heavy Rain + Flooded Road & Night without Streetlights & Dense Fog + Construction Area  & Strong Wind + Loose Dry Surface & Setting Sun + Slippery Roads & Sunset + Worn Road Markings \\
\midrule
\textbf{Description} & Intense rainfall causing water accumulation on roads, affecting vehicle traction and visibility. & Insufficient lighting makes it difficult for drivers to see road conditions, especially in areas without streetlights. & Low visibility making it difficult for drivers to detect construction areas ahead, increasing collision risk. & Strong winds may lift dust from the road, reducing visibility and road friction. Crosswinds can also cause vehicle deviation. & Insufficient light and unexpected obstacles and temporary signs in construction areas. & Oncoming headlights at sunset and worn road markings may lead to driver judgment errors. \\
\midrule
\textbf{Figure} & \begin{minipage}[c]{2cm}\centering\includegraphics[width=2cm]{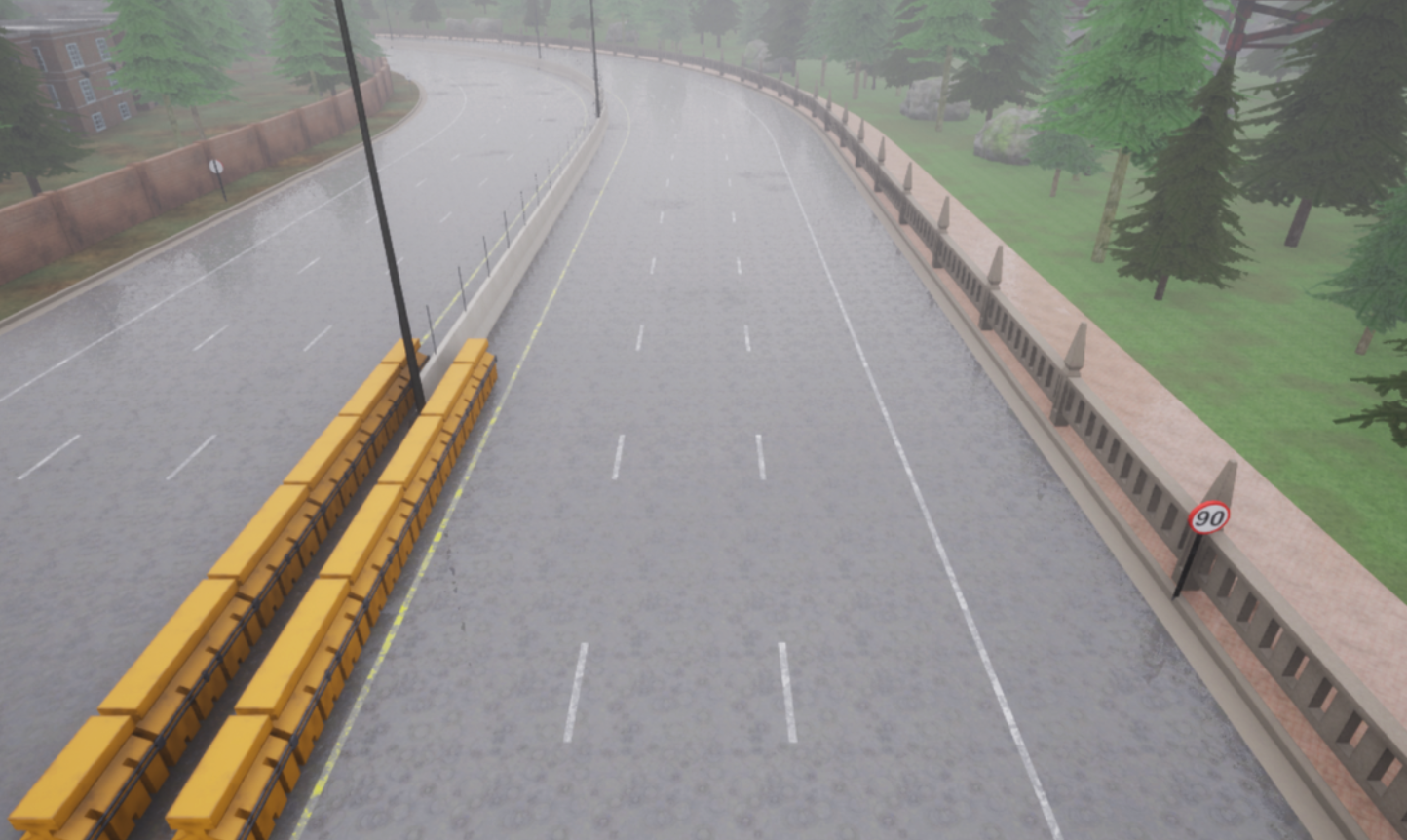}\end{minipage} & \begin{minipage}[c]{2cm}\centering\includegraphics[width=2cm]{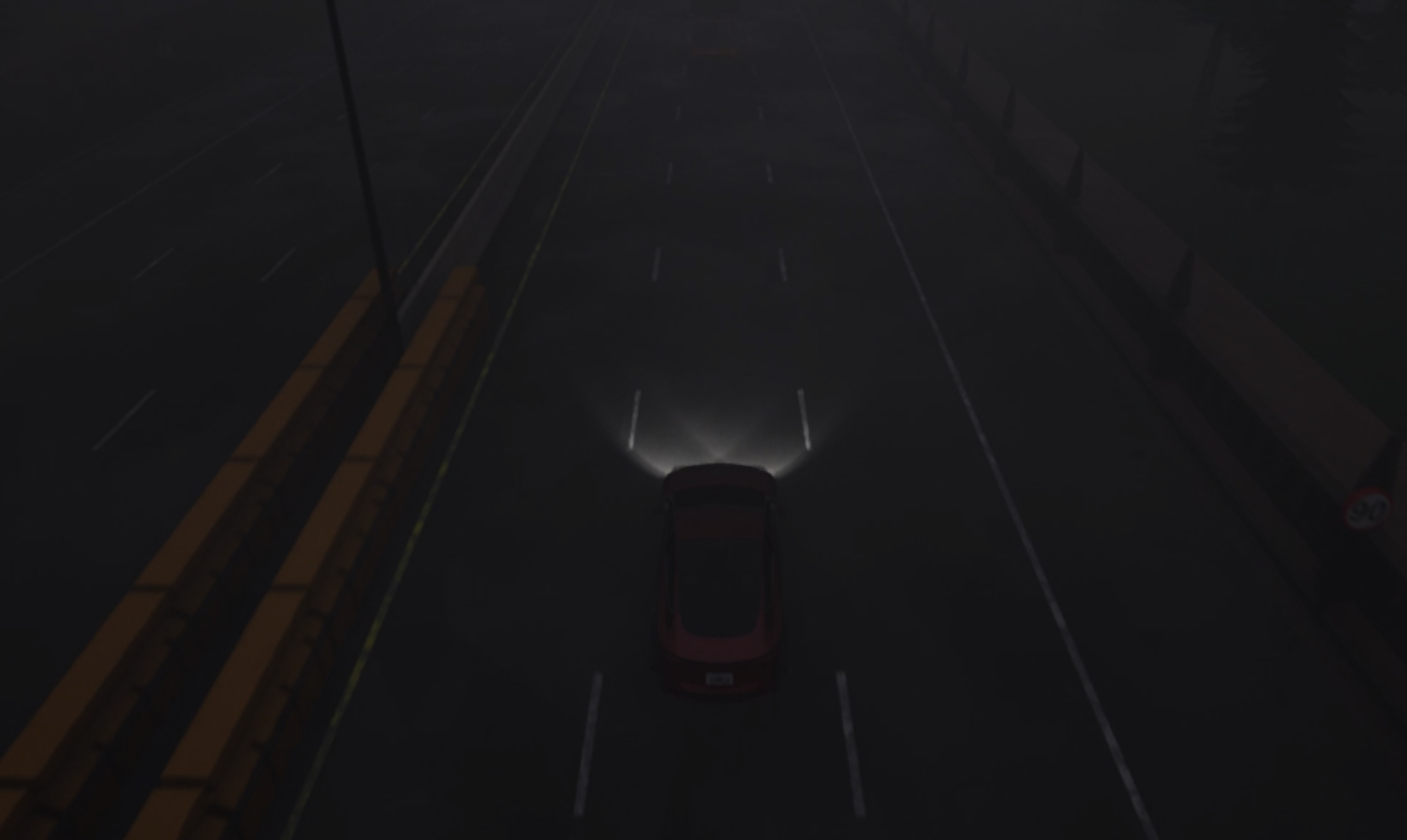}\end{minipage} & \begin{minipage}[c]{2cm}\centering\includegraphics[width=2cm]{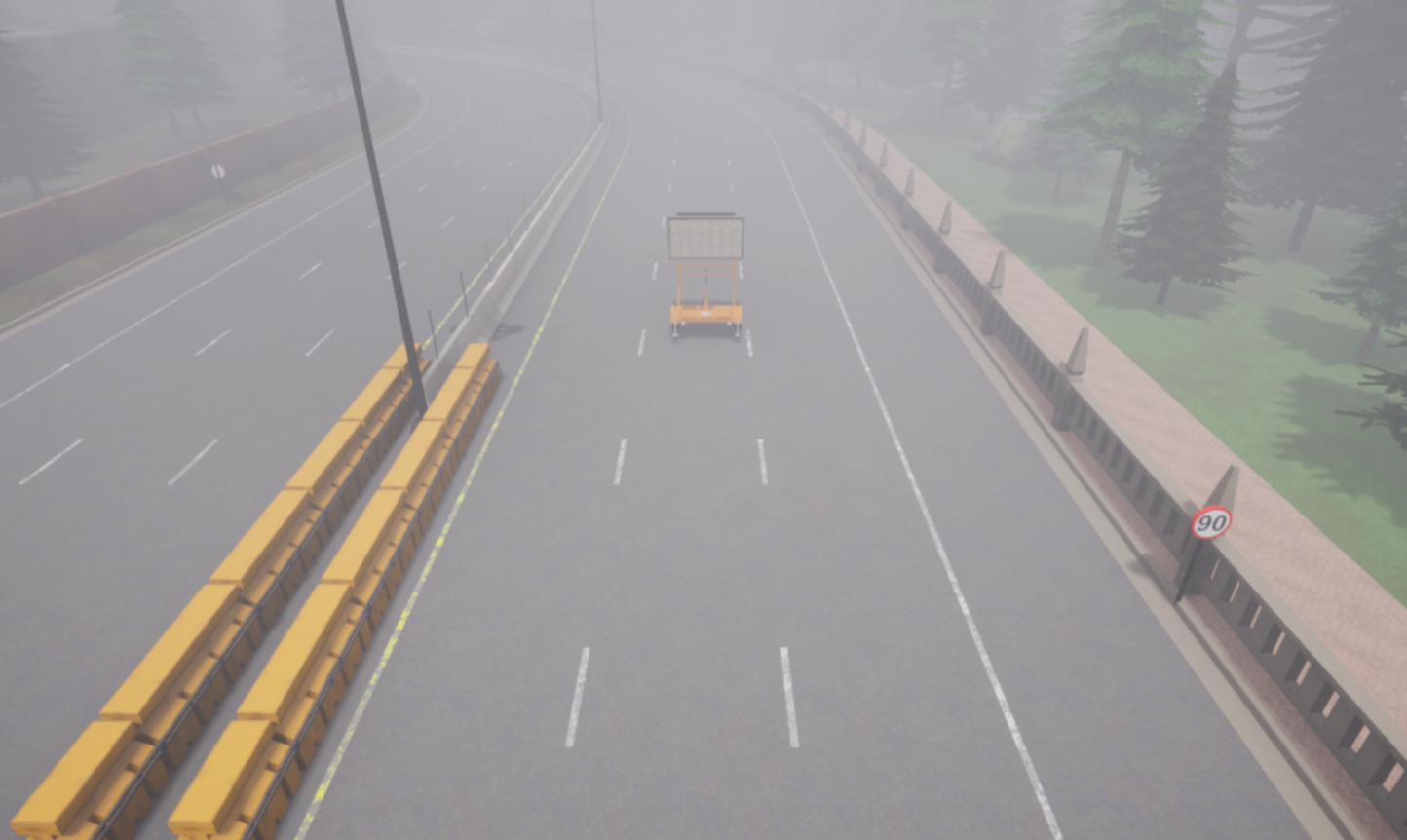}\end{minipage} & \begin{minipage}[c]{2cm}\centering\includegraphics[width=2cm]{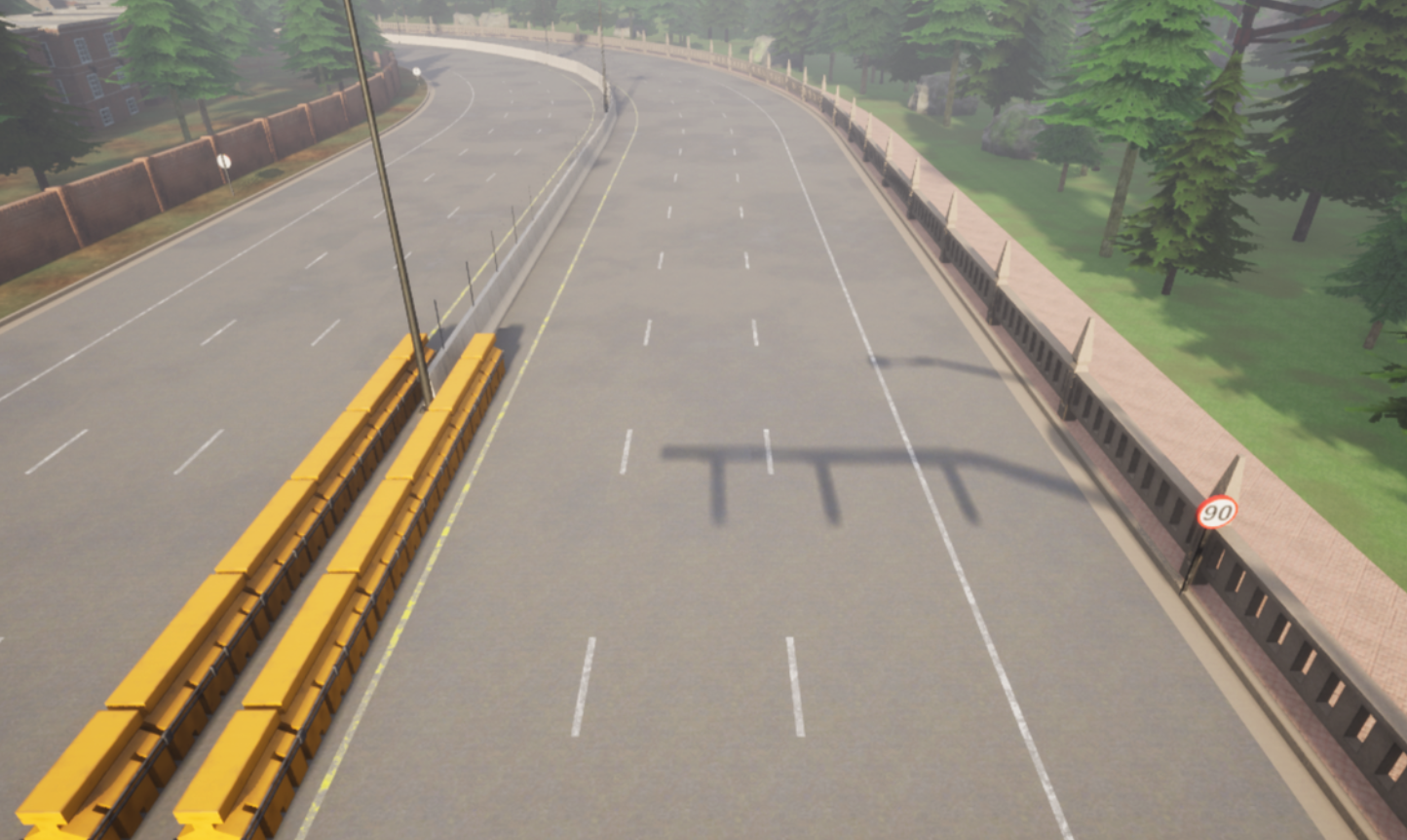}\end{minipage} & \begin{minipage}[c]{2cm}\centering\includegraphics[width=2cm]{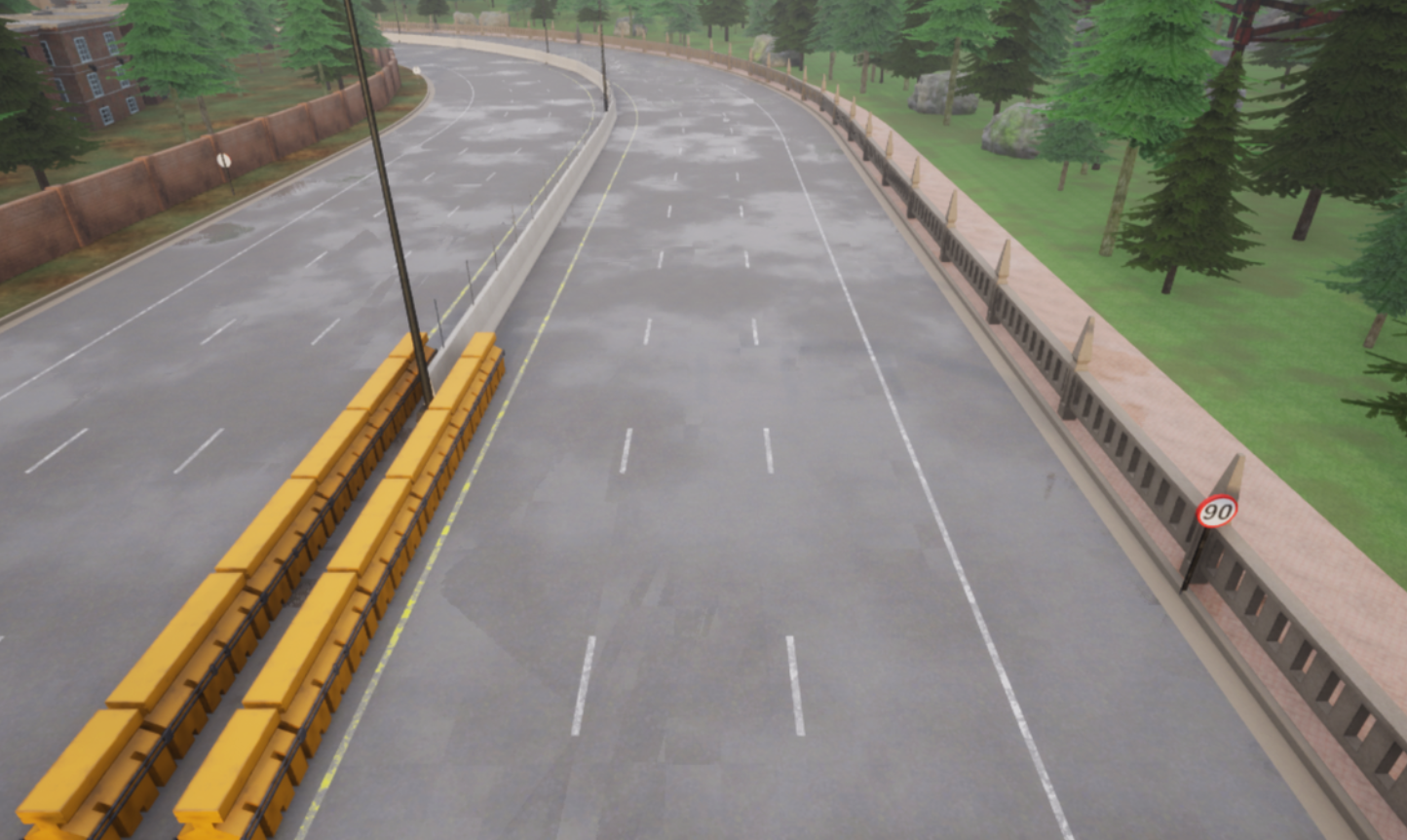}\end{minipage} & \begin{minipage}[c]{2cm}\centering\includegraphics[width=2cm]{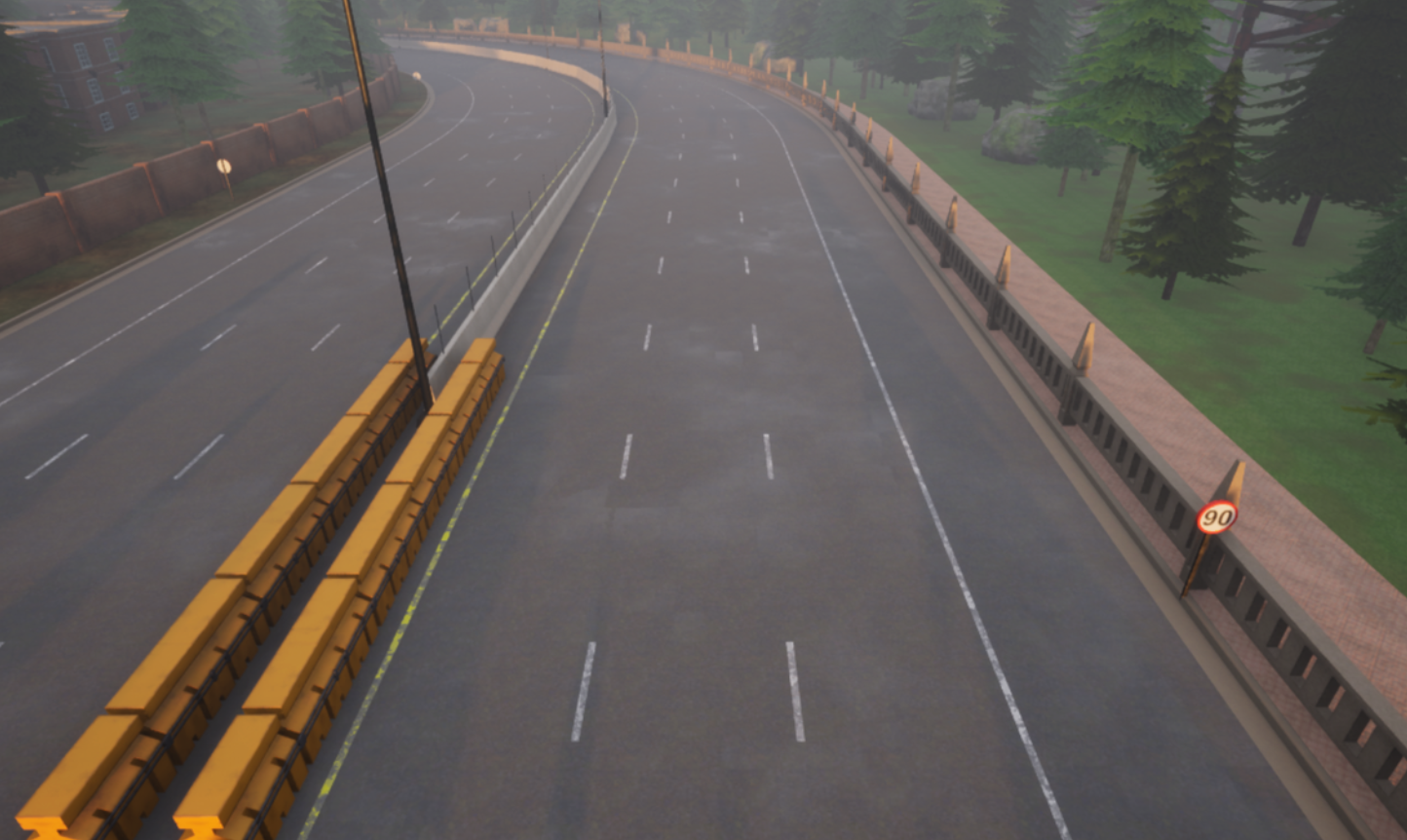}\end{minipage} \\
\bottomrule
\end{tabular}
\end{small}
}
\label{tb:static}
\end{table*}

The risk scenario generation process using the CBN, as detailed in Alg.~\ref{alg:gene}, involves a series of steps. Initially, the algorithm replicates the current risk scenario for each static element combination (line 3). 
Then it checks if the current scenario satisfies the precondition for that combination (line 4). 
If it does, the scenario is set according to the combination's specifics (line 5).
Next, the algorithm, utilizing CBN, deduces a set of potential risky actions (line 6). 
For each identified action, if the current scenario satisfies the action's precondition, the algorithm sets the action in the scenario, calculates its Severity ($S$) and Exposure ($E$), and computes the Risk Priority ($RP$) (line 7-11). These values are then used to save the scenario and its $RP$ in the collection of risk scenarios. Finally, the algorithm compiles and returns this collection of risk scenarios (line 12).

In essence, this algorithm assesses risks in functional scenarios by analyzing static element combinations and risky actions, generating a collection of scenarios ranked by their risk priority.

\begin{algorithm}
\caption{Risk Scenario Generation Based on CBN}
\begin{algorithmic}[1]
\REQUIRE $originalScenario$: functional scenario description
\ENSURE  $riskScenarios$: a collection of generated risk scenarios

\STATE $riskScenarios \leftarrow \emptyset$

\FORALL{$staticCombination$}
    \STATE $currScenario \leftarrow \text{copy}(originalScenario)$
    \IF{$\text{precond}(currScenario, staticCombination)$}
        \STATE $\text{set}(currScenario, staticCombination)$
        \STATE $riskActions \leftarrow \text{deduce}(currScenario, CBN)$
        \FORALL{$action \in riskActions$}
            \IF{$\text{precond}(currScenario, action)$}
                \STATE $\text{setAction}(currScenario, action)$
                \STATE $S, E \leftarrow \text{calcSE}(currScenario)$
                \STATE $RP \leftarrow \text{calcRP}(S, E)$
                \STATE $riskScenarios \leftarrow riskScenarios \cup \{(currScenario, RP)\}$
            \ENDIF
        \ENDFOR
    \ENDIF
\ENDFOR

\RETURN $riskScenarios$
\end{algorithmic}
\label{alg:gene}
\end{algorithm}

\subsection{Test execution}
\label{chap:3.4}
The testing execution module is built based on the CARLA\cite{carla} simulation environment. Since the scenarios generated by Alg.~\ref{alg:gene} are high-level functional scenario descriptions, they cannot be directly imported to CARLA and executed. Therefore, step-by-step conversion is required. The conversion process is inspired by the method proposed in \cite{menzelFunctionalLogicalScenarios2019}:

Firstly, the functional scenario description is converted into a logical scenario representation. In this representation, the scenario elements are replaced by variables, parameter ranges, and environmental effects instead of concrete parameters. Then, through the scenario editor, the logical scenario is mapped to an executable concrete scenario. The process includes randomization of scenario parameters, insertion of traffic flows and events, etc. Finally, it can be loaded and run directly in CARLA.

Based on the executable scenarios, the testing execution module initiates a pre-trained ADS, which encompasses an image recognition perception model based on ResNet\cite{resnet}, and an end-to-end decision and planning model trained with Deep Deterministic Policy Gradients (DDPG)\cite{ddpg}. The use of DDPG, which directly determines the next state of the ADS, allows us to bypass traditional Proportion-Integration-Differentiation (PID) control methods. This ADS serves as the test object, playing a critical role in our evaluation framework.

Furthermore, the testing execution module includes a misbehavior detector designed to analyze and assess the performance of the ADS in real time within the CARLA simulation environment. It evaluates driving quality and the hazard level of scenarios based on three key aspects: \textbf{safety, comfort, and efficiency}, as detailed in Table \ref{tb:rules}. 

In terms of \textbf{safety}, the misbehavior detector focuses on monitoring collision-related situations, such as minimum time-to-collision (TTC) with pedestrians or other vehicles, collision counts, and the frequency of violations of traffic signals and prohibitory signs. These metrics are crucial in assessing the hazard level of each scenario, serving as key indicators for determining potential risks. They not only help identify immediate risks but also provide vital data support for long-term safety performance analysis. By closely monitoring these safety-related metrics, we can effectively evaluate the severity and frequency of hazardous events within each scenario, making them integral to our comprehensive assessment of scenario danger.

Regarding \textbf{comfort}, this module pays attention to metrics such as changes in vehicle acceleration, speed fluctuations, sudden changes in steering angles, and lateral deviation from the center line. These metrics help assess the passenger experience, ensure passenger comfort, and indirectly reflect the stability of driving behavior.

For \textbf{efficiency} evaluation, the misbehavior detector records metrics such as average speed, lane changes, brake activation, and reaction times to evaluate the vehicle's driving efficiency and adaptability. These metrics not only reflect the ADS's ability to handle complex road conditions but also reveal its advantages in terms of efficiency and resource utilization.

Through the recording and analysis of the above-mentioned metrics, we can comprehensively assess the driving quality of the ADS and the associated risk level of each scenario. 

\begin{table}[!ht]
\caption{Evaluation metrics for misbehavior}
\label{tb:rules}
\begin{tabularx}{\columnwidth}{>{\hsize=1\hsize}X|>{\hsize=12 \hsize}X|>{\hsize=7\hsize}X}
\toprule
\textbf{Aspect}  & \textbf{Metric} & \textbf{Category} \\ \midrule
\multicolumn{1}{c|}{\multirow{6}{*}{Safety}}     & Distance, Time, Crash                                                                                        & Generic                  \\
\multicolumn{1}{c|}{}                            & Mean Speed, Speed Standard Deviation                                                                                         & Speed and Acceleration    \\
\multicolumn{1}{c|}{}                            & Lateral Position Standard Deviation                                                                                           & Lateral Position         \\
\multicolumn{1}{c|}{}                            & Lateral Acceleration, Trajectory Offset                                                                      & Lateral Speed       \\
\multicolumn{1}{c|}{}                            & Collision, Time-To-Collision, Collision Count                                                                & Headway Position         \\
\multicolumn{1}{c|}{}                            & Traffic Sign Compliance                                                                                      & Traffic Signs            \\ \midrule
\multicolumn{1}{c|}{\multirow{3}{*}{Comfort}}    & Throttle Pedal Position, Throttle Pedal Speed                                                                & Throttle Pedal           \\
\multicolumn{1}{c|}{}                            & Lane Change Count, Lane Change Time                                                                          & Lane Change         \\
\multicolumn{1}{c|}{}                            & Steering Angle, Steering Angle SD, Steering Angle Reversal Rate, Steering   Angle Speed                     & Steering                 \\ \midrule
\multicolumn{1}{c|}{\multirow{3}{*}{Efficiency}} & Mean Speed                                                                                                   & Speed and Acceleration    \\
\multicolumn{1}{c|}{}                            & Lane Change Count, Lane Change Time                                                                          & Lane Change         \\
\multicolumn{1}{c|}{}                            & Braking Count, Reaction Time                                                                              & Braking                  \\ \bottomrule
\end{tabularx}
\end{table}

\section{Evaluation}
Our experiments are conducted in an ADS simulation environment powered by CARLA 0.9.6. It incorporates an ADS with perception, decision-making, and control modules. The hardware platform consists of an Intel i7-8700K CPU, Nvidia RTX 3080 Ti GPU, and 32GB RAM to ensure efficient execution.

The ADS includes a ResNet-based image recognition perception model for detecting surrounding objects, and an end-to-end decision \& planning model trained via Deep Deterministic Policy Gradients (DDPG) to directly map perceptions to driving actions. Furthermore, the comprehensive misbehavior detector introduced in Section \ref{chap:3.4} is implemented to assess the ego vehicle's driving performance from dimensions of safety, comfort, and efficiency during runtime. It records rich metrics on collisions, traffic violations, acceleration changes, etc.

The simulation contains a diverse set of 5 functional scenario types covering major capabilities like lane keeping, obstacle avoidance, car following, etc. The standardized yet representative experimentation environment allows a reliable assessment of our hazard scenario generation techniques.

To empirically assess the framework presented in this article, we delve into the following research questions (RQs):

\begin{itemize}
    \item \textbf{RQ1:} To what extent can the CBN-based Scenario Generation framework generate high-risk driving scenarios?
    \item \textbf{RQ2:} Compared to existing approaches, does our method generate risk scenarios more effectively?
    \item \textbf{RQ3:} What is the run-time efficiency of our proposed approach?
\end{itemize}

\subsection{RQ1: Risk scenarios generated}
To evaluate the effectiveness of CBN-based scenario generation (RQ1), we implemented the scenario generation module in the framework based on Alg.~\ref{alg:gene}. Specifically, we present 5 typical autonomous driving scenarios as seeds, including lane keeping, pedestrian crossing, following, overtaking, and turning left at intersections. These scenarios cover typical autonomous driving functions such as lane keeping, avoiding obstacles, following, overtaking, and turning.

\begin{figure}[htbp]
\centerline{\includegraphics[width=0.5\textwidth]{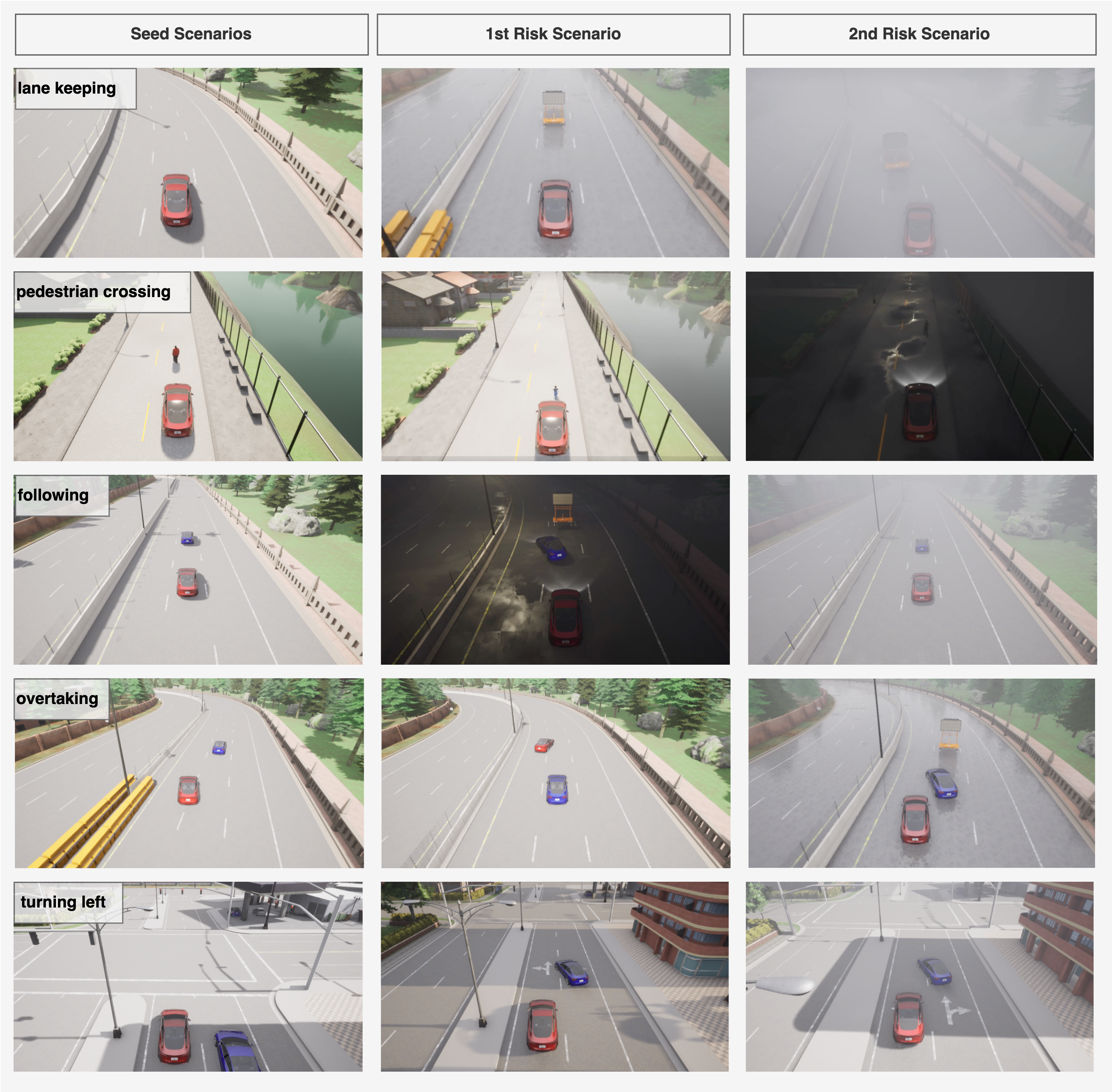}}
\caption{Risk scenarios generated with CBN}
\label{fig:case}
\end{figure}

On the basis of these seed scenarios, the scenario generation module iterates through possible combinations of static elements, including weather, road conditions, lighting, etc. Meanwhile, it infers potentially high-risk actions under these combinations based on the CBN, such as pedestrians suddenly crossing or front vehicles emergency braking. The module gradually checks the precondition of the current scenario and sets subsequent actions following Alg.~\ref{alg:gene}, eventually generating 89 risk scenarios.

\begin{figure}[htbp]
\centerline{\includegraphics[width=0.5\textwidth]{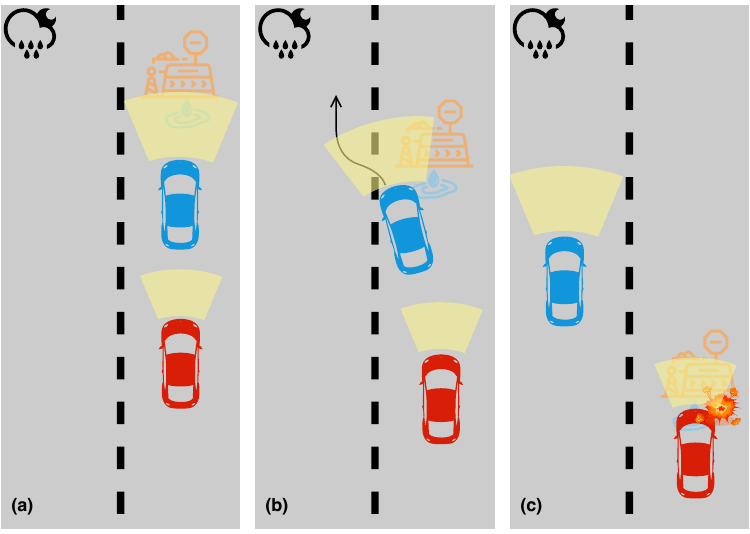}}
\caption{Case study: risk scenario generated from the following scenario}
\label{fig:case_study}
\end{figure}

We rendered these 5 seed scenarios and their corresponding generated risk scenarios within the CARLA simulation environment. Fig.~\ref{fig:case} shows the original seed scenarios and some examples of generated risk scenarios. The first column in Fig.~\ref{fig:case} shows the original seed scenarios, while the second and third columns present the risk scenarios generated based on our method. For instance, in the following scenario depicted in the third row, the original scenario involves an ego car (red car) maintaining a safe distance behind the background vehicle (blue car) under favorable weather conditions. Fig.~\ref{fig:case_study} illustrates the process of collision in the first generated risk scenario. At the outset, the ego vehicle (red car) follows the background vehicle (blue car) during a rainy night without streetlight illumination (as shown in Fig.\ref{fig:case_study}-a). Subsequently, the background vehicle detects an obstacle ahead and shifts to the left lane to avoid it. Meanwhile, unaware of the road conditions ahead, the ego vehicle continues to maintain its original course (as depicted in Fig.\ref{fig:case_study}-b). Finally, due to insufficient lighting and slippery road conditions, the ego vehicle fails to timely avoid the obstacle, leading to a collision (as illustrated in Fig.~\ref{fig:case_study}-c).

\begin{table}[htbp]
\centering
\begin{tabular}{|p{0.45\textwidth}|}
\hline
\textbf{Finding 1: } The implementation of our CBN-based framework has effectively resulted in the generation of an average of 17 high-risk driving scenarios for each initial seed scenario. This significant yield highlights the framework's ability to not only identify crucial risk factors from historical accident data but also to creatively incorporate these factors into realistic and dynamic scenarios. \\ \hline
\end{tabular}%
\end{table}

\subsection{RQ2: Efficiency of CBN-based scenario generation}
To demonstrate the efficiency of our CBN-based scenario generation method (RQ2), we compare it with two baseline methods: Random and DriveFuzz \cite{kimDriveFuzzDiscoveringAutonomous2022}. The Random method randomly adds obstacles and alters weather and road conditions to produce scenarios. However, the resulting hazardous factors may ignore traffic rules and other road participants. DriveFuzz modifies driving tasks, weather, and road defects based on driving quality. However, the iteration process is time-consuming.

In contrast, our method identifies risk variables and actions from real-world accident data. The search for scenario generation is efficient and targeted.

\begin{figure}[htbp]
\centerline{\includegraphics[width=0.5\textwidth]{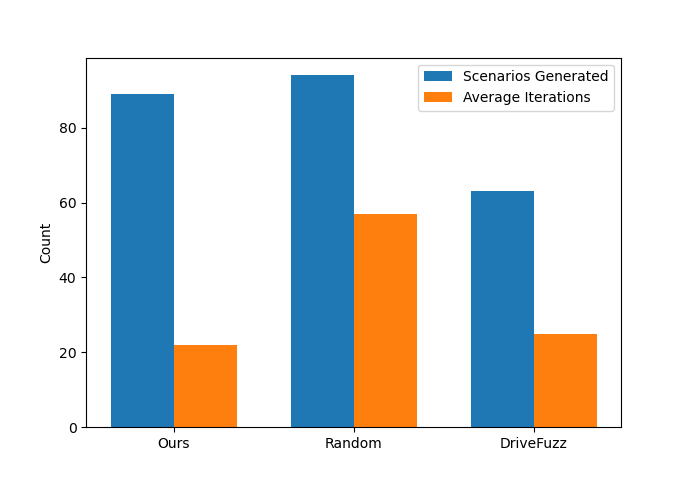}}
\caption{Comparison of scenario generation methods}
\label{fig:res_sce}
\end{figure}

As shown in Fig.~\ref{fig:res_sce}, we derive 89 risk scenarios from 5 seed cases using our approach within 22 iterations on average, which demonstrates a remarkable efficiency, generating one risk scenario on average every 1.2 iterations. This performance is notably superior to the Random and DriveFuzz methods, which required an average of 3.1 and 1.98 iterations respectively to generate a single risk scenario. 
The results prove that leveraging CBN can largely improve the efficiency of risk scenario generation for ADS.

\begin{table}[htbp]
\centering
\begin{tabular}{|p{0.45\textwidth}|}
\hline
\textbf{Finding 2: } In comparing our CBN-based scenario generation method with the traditional Random approach and the DriveFuzz methodology, we observed a marked improvement in the efficiency of generating risk scenarios. On average, our approach required 1.9 fewer iterations compared to the Random method and 0.8 fewer iterations than the DriveFuzz method. 
\\ \hline
\end{tabular}%
\end{table}

\subsection{RQ3: Runtime efficiency of CBN-based scenario generation}
To evaluate the runtime efficiency of our approach (RQ3), we record the time consumption of the overall pipeline. The construction of CBN takes 5 hours since it requires accident data preprocessing and causal discovery from scratch.

However, this process can be highly paralleled, as the generation for different scenarios is independent. Once built, the model is reusable for various cases. 
Table \ref{tab:performance} illustrates the performance of three different algorithms during the scenario generation phase. We recorded the total time spent on scenario generation and the number of times the simulator was initiated. Comparative analysis revealed that our method significantly reduces the time required for scenario generation. On average, each effective high-risk scenario was generated in 30 seconds, which is a reduction of 35 seconds and 60 seconds compared to the DriveFuzz and Random methods, respectively. Furthermore, the reduced number of simulator initiations, as observed in our method, significantly enhances runtime performance by minimizing the frequency of simulator startups.

\begin{table}[htbp]
\caption{Runtime Performance}
\label{tab:performance}
\begin{tabular}{@{}lcc@{}}
\toprule
Method        & \multicolumn{1}{l}{Total Time (mins)} & \multicolumn{1}{l}{Simulator Start-up Times} \\ \midrule
Random        & 142                                   & 303                                          \\
DriveFuzz     & 62                                    & 121                                          \\
\textbf{Ours} & \textbf{45}                           & \textbf{107}                                 \\ \bottomrule
\end{tabular}
\end{table}

In conclusion, the offline construction of CBN pays a one-time cost for scenario generation acceleration. Scenario generation driven by CBN demonstrates higher time efficiency compared with baselines. 
\begin{table}[htbp]
\centering
\begin{tabular}{|p{0.45\textwidth}|}
\hline
\textbf{Finding 3: } Our study's runtime efficiency analysis indicates that, while the initial setup and construction of the CBN require a significant investment of time, the subsequent scenario generation process is notably faster and more efficient than existing methods. The reason lies in the directed generation simplified by pre-constructed CBN, avoiding many redundant iterations.  \\ \hline
\end{tabular}%
\end{table}

\section{Threats to Validity}

This section discusses the internal and external threats to the validity of our study, aiming to provide a comprehensive understanding of the potential limitations and implications of our research.

\subsection{Internal threats}
Internal threats to validity in our study primarily arise from the methodological choices, experimental design, and data sources used. These include:

\begin{itemize}
    \item \textbf{Data Reliability and Bias:} The CBN relies on the Maryland accident data. The inherent biases or inaccuracies in this dataset may impact the model's accuracy and reliability.
    \item \textbf{Modeling Assumptions and Simplifications:} Assumptions made during the construction of the CBN, including the selection of variables, the establishment of causal relationships, and simplifications for computational feasibility, may not fully capture the complexity of real-world driving scenarios.
    \item \textbf{Simulation Limitations:} Our reliance on the CARLA simulation environment for testing scenarios could limit the representation of real-world dynamics. Simulations may not fully encompass the unpredictability and variety of real-world conditions.
    \item \textbf{Algorithmic Constraints:} The design and implementation of algorithms for scenario generation and risk assessment may introduce limitations due to computational constraints, potentially affecting the breadth and depth of scenario coverage.
\end{itemize}

\subsection{External threats}
External threats pertain to the generalizability and applicability of our findings to broader contexts, including different systems, environments, and conditions. These include:

\begin{itemize}
    \item \textbf{Generalizability Across Systems:} The scenarios and findings are based on a specific ADS setup within the CARLA simulation environment. The transferability of our results to other ADS, software architectures, and hardware configurations remains uncertain.
    \item \textbf{Scalability and Performance:} The scalability of our approach in larger and more diverse datasets, including real-world data, has not been fully tested. The performance in different and evolving technological contexts is also a concern.
    \item \textbf{Evolution of Autonomous Driving Technologies:} The rapidly evolving nature of autonomous driving technologies and regulatory environments could lead to new risk factors, scenarios, and testing requirements not accounted for in our current framework.
    \item \textbf{Diversity of Driving Conditions:} Our study may not adequately cover the wide range of geographical, climatic, and infrastructural variations that affect driving scenarios globally.
\end{itemize}

In conclusion, while our study contributes valuable insights into CBN-based scenario generation for ADS testing, these internal and external threats highlight the need for cautious interpretation and suggest areas for future research and development.

\section{Discussion and Conclusion }
This paper proposes a CBN-based scenario generation framework for ADS testing using Causal Bayesian Networks. Compared to prior arts, our solution demonstrates higher efficiency and accuracy in discovering critical test cases. The probabilistic causal modeling offers a comprehensive perspective on accident causation. The inferred risk knowledge further directs the efficient generation of test scenarios. Experiments prove the capability of our system to generate risk scenarios with lower iterations.

Despite the strengths of our approach, it also presents opportunities for enhancement:

\begin{itemize}
    \item Despite incorporating prior knowledge, the construction of the CBN still depends on the quality of accident data. A key area for further exploration is how to integrate multiple accident databases to more comprehensively depict driving scenarios.
    \item While the iteration process is effective, the escalating complexity of the CBN model may lead to increased time demands in both its construction and inferential processes.
\end{itemize}

In conclusion, we present a virtual testing framework focused on scenario generation. A causal Bayesian Network encodes probabilistic risk knowledge, enabling the automatic generation of safety-critical driving scenarios. Experiments demonstrate complex yet physically realistic risk situations can be generated within fewer iterations. 
Future efforts will be directed towards enhancing the efficiency of causal modeling and optimization of search algorithms. Additionally, comprehensive case analyses will be conducted to further validate and refine our approach.

\textbf{\bibliographystyle{ACM-Reference-Format}
\bibliography{sample-base}}


\begin{thebibliography}{44}


\ifx \showCODEN    \undefined \def \showCODEN     #1{\unskip}     \fi
\ifx \showDOI      \undefined \def \showDOI       #1{#1}\fi
\ifx \showISBNx    \undefined \def \showISBNx     #1{\unskip}     \fi
\ifx \showISBNxiii \undefined \def \showISBNxiii  #1{\unskip}     \fi
\ifx \showISSN     \undefined \def \showISSN      #1{\unskip}     \fi
\ifx \showLCCN     \undefined \def \showLCCN      #1{\unskip}     \fi
\ifx \shownote     \undefined \def \shownote      #1{#1}          \fi
\ifx \showarticletitle \undefined \def \showarticletitle #1{#1}   \fi
\ifx \showURL      \undefined \def \showURL       {\relax}        \fi
\providecommand\bibfield[2]{#2}
\providecommand\bibinfo[2]{#2}
\providecommand\natexlab[1]{#1}
\providecommand\showeprint[2][]{arXiv:#2}

\bibitem[mar(2024)]%
        {marylandBenchmarkReports}
 \bibinfo{year}{2024}\natexlab{}.
\newblock \bibinfo{title}{{B}enchmark {R}eports - {P}ages --- mva.maryland.gov}.
\newblock \bibinfo{howpublished}{\url{https://mva.maryland.gov/safety/Pages/mhso/benchmark-reports.aspx}}.
\newblock
\newblock
\shownote{[Accessed 11-01-2024]}.


\bibitem[dat(2024)]%
        {data_catalog}
 \bibinfo{year}{2024}\natexlab{}.
\newblock \bibinfo{title}{{C}rash {D}ata - {C}atalog --- catalog.data.gov}.
\newblock \bibinfo{howpublished}{\url{https://catalog.data.gov/dataset/crash-data}}.
\newblock
\newblock
\shownote{[Accessed 11-01-2024]}.


\bibitem[cal(2024)]%
        {calderdaleHomeCalderdale}
 \bibinfo{year}{2024}\natexlab{}.
\newblock \bibinfo{title}{{H}ome | {C}alderdale {D}ata {W}orks --- dataworks.calderdale.gov.uk}.
\newblock \bibinfo{howpublished}{\url{https://dataworks.calderdale.gov.uk/}}.
\newblock
\newblock
\shownote{[Accessed 11-01-2024]}.


\bibitem[Bagschik et~al\mbox{.}(2018)]%
        {ontologygeneration}
\bibfield{author}{\bibinfo{person}{Gerrit Bagschik}, \bibinfo{person}{Till Menzel}, {and} \bibinfo{person}{Markus Maurer}.} \bibinfo{year}{2018}\natexlab{}.
\newblock \showarticletitle{Ontology based Scene Creation for the Development of Automated Vehicles}. In \bibinfo{booktitle}{\emph{2018 IEEE Intelligent Vehicles Symposium (IV)}}. \bibinfo{pages}{1813--1820}.
\newblock
\urldef\tempurl%
\url{https://doi.org/10.1109/IVS.2018.8500632}
\showDOI{\tempurl}


\bibitem[Baidu(2024)]%
        {baiduApollox5F00x53D1x8005x793Ex533A}
\bibfield{author}{\bibinfo{person}{Baidu}.} \bibinfo{year}{2024}\natexlab{}.
\newblock \bibinfo{title}{{A}pollo; --- apollo.baidu.com}.
\newblock \bibinfo{howpublished}{\url{https://apollo.baidu.com/}}.
\newblock
\newblock
\shownote{[Accessed 11-01-2024]}.


\bibitem[Bl{\"o}baum et~al\mbox{.}(2022)]%
        {dowhy_gcm}
\bibfield{author}{\bibinfo{person}{Patrick Bl{\"o}baum}, \bibinfo{person}{Peter G{\"o}tz}, \bibinfo{person}{Kailash Budhathoki}, \bibinfo{person}{Atalanti~A. Mastakouri}, {and} \bibinfo{person}{Dominik Janzing}.} \bibinfo{year}{2022}\natexlab{}.
\newblock \showarticletitle{DoWhy-GCM: An extension of DoWhy for causal inference in graphical causal models}.
\newblock \bibinfo{journal}{\emph{arXiv preprint arXiv:2206.06821}} (\bibinfo{year}{2022}).
\newblock


\bibitem[Calò et~al\mbox{.}(2020)]%
        {plan_avoid}
\bibfield{author}{\bibinfo{person}{Alessandro Calò}, \bibinfo{person}{Paolo Arcaini}, \bibinfo{person}{Shaukat Ali}, \bibinfo{person}{Florian Hauer}, {and} \bibinfo{person}{Fuyuki Ishikawa}.} \bibinfo{year}{2020}\natexlab{}.
\newblock \showarticletitle{Generating Avoidable Collision Scenarios for Testing Autonomous Driving Systems}. In \bibinfo{booktitle}{\emph{2020 IEEE 13th International Conference on Software Testing, Validation and Verification (ICST)}}. \bibinfo{pages}{375--386}.
\newblock
\urldef\tempurl%
\url{https://doi.org/10.1109/ICST46399.2020.00045}
\showDOI{\tempurl}


\bibitem[Chia et~al\mbox{.}(2022)]%
        {chiaRiskAssessmentMethodologies2022}
\bibfield{author}{\bibinfo{person}{Wei Ming~Dan Chia}, \bibinfo{person}{Sye~Loong Keoh}, \bibinfo{person}{Cindy Goh}, {and} \bibinfo{person}{Christopher Johnson}.} \bibinfo{year}{2022}\natexlab{}.
\newblock \showarticletitle{Risk {{Assessment Methodologies}} for {{Autonomous Driving}}: {{A Survey}}}.
\newblock \bibinfo{journal}{\emph{IEEE Transactions on Intelligent Transportation Systems}} \bibinfo{volume}{23}, \bibinfo{number}{10} (\bibinfo{date}{Oct.} \bibinfo{year}{2022}), \bibinfo{pages}{16923--16939}.
\newblock
\showISSN{1558-0016}
\urldef\tempurl%
\url{https://doi.org/10.1109/TITS.2022.3163747}
\showDOI{\tempurl}


\bibitem[de~Gelder et~al\mbox{.}(2022)]%
        {scenariosdefine}
\bibfield{author}{\bibinfo{person}{Erwin de Gelder}, \bibinfo{person}{Jan-Pieter Paardekooper}, \bibinfo{person}{Arash~Khabbaz Saberi}, \bibinfo{person}{Hala Elrofai}, \bibinfo{person}{Olaf Op~den Camp}, \bibinfo{person}{Steven Kraines}, \bibinfo{person}{Jeroen Ploeg}, {and} \bibinfo{person}{Bart De~Schutter}.} \bibinfo{year}{2022}\natexlab{}.
\newblock \showarticletitle{Towards an Ontology for Scenario Definition for the Assessment of Automated Vehicles: An Object-Oriented Framework}.
\newblock \bibinfo{journal}{\emph{IEEE Transactions on Intelligent Vehicles}} \bibinfo{volume}{7}, \bibinfo{number}{2} (\bibinfo{year}{2022}), \bibinfo{pages}{300--314}.
\newblock
\urldef\tempurl%
\url{https://doi.org/10.1109/TIV.2022.3144803}
\showDOI{\tempurl}


\bibitem[Ding et~al\mbox{.}(2023)]%
        {dingSurveySafetyCriticalDriving2023}
\bibfield{author}{\bibinfo{person}{Wenhao Ding}, \bibinfo{person}{Chejian Xu}, \bibinfo{person}{Mansur Arief}, \bibinfo{person}{Haohong Lin}, \bibinfo{person}{Bo Li}, {and} \bibinfo{person}{Ding Zhao}.} \bibinfo{year}{2023}\natexlab{}.
\newblock \showarticletitle{A {{Survey}} on {{Safety-Critical Driving Scenario Generation}}{\textemdash}{{A Methodological Perspective}}}.
\newblock \bibinfo{journal}{\emph{IEEE Transactions on Intelligent Transportation Systems}} \bibinfo{volume}{24}, \bibinfo{number}{7} (\bibinfo{date}{July} \bibinfo{year}{2023}), \bibinfo{pages}{6971--6988}.
\newblock
\showISSN{1558-0016}
\urldef\tempurl%
\url{https://doi.org/10.1109/TITS.2023.3259322}
\showDOI{\tempurl}


\bibitem[Dosovitskiy et~al\mbox{.}(2017)]%
        {carla}
\bibfield{author}{\bibinfo{person}{Alexey Dosovitskiy}, \bibinfo{person}{German Ros}, \bibinfo{person}{Felipe Codevilla}, \bibinfo{person}{Antonio Lopez}, {and} \bibinfo{person}{Vladlen Koltun}.} \bibinfo{year}{2017}\natexlab{}.
\newblock \showarticletitle{{CARLA}: {An} Open Urban Driving Simulator}. In \bibinfo{booktitle}{\emph{Proceedings of the 1st Annual Conference on Robot Learning}} \emph{(\bibinfo{series}{Proceedings of Machine Learning Research}, Vol.~\bibinfo{volume}{78})}, \bibfield{editor}{\bibinfo{person}{Sergey Levine}, \bibinfo{person}{Vincent Vanhoucke}, {and} \bibinfo{person}{Ken Goldberg}} (Eds.). \bibinfo{publisher}{PMLR}, \bibinfo{pages}{1--16}.
\newblock
\urldef\tempurl%
\url{https://proceedings.mlr.press/v78/dosovitskiy17a.html}
\showURL{%
\tempurl}


\bibitem[Feng et~al\mbox{.}(2021)]%
        {fengTestingScenarioLibrary2021}
\bibfield{author}{\bibinfo{person}{Shuo Feng}, \bibinfo{person}{Yiheng Feng}, \bibinfo{person}{Chunhui Yu}, \bibinfo{person}{Yi Zhang}, {and} \bibinfo{person}{Henry~X. Liu}.} \bibinfo{year}{2021}\natexlab{}.
\newblock \showarticletitle{Testing {{Scenario Library Generation}} for {{Connected}} and {{Automated Vehicles}}, {{Part I}}: {{Methodology}}}.
\newblock \bibinfo{journal}{\emph{IEEE Transactions on Intelligent Transportation Systems}} \bibinfo{volume}{22}, \bibinfo{number}{3} (\bibinfo{date}{March} \bibinfo{year}{2021}), \bibinfo{pages}{1573--1582}.
\newblock
\showISSN{1558-0016}
\urldef\tempurl%
\url{https://doi.org/10.1109/TITS.2020.2972211}
\showDOI{\tempurl}


\bibitem[Feng et~al\mbox{.}(2023)]%
        {fengDenseReinforcementLearning2023}
\bibfield{author}{\bibinfo{person}{Shuo Feng}, \bibinfo{person}{Haowei Sun}, \bibinfo{person}{Xintao Yan}, \bibinfo{person}{Haojie Zhu}, \bibinfo{person}{Zhengxia Zou}, \bibinfo{person}{Shengyin Shen}, {and} \bibinfo{person}{Henry~X. Liu}.} \bibinfo{year}{2023}\natexlab{}.
\newblock \showarticletitle{Dense Reinforcement Learning for Safety Validation of Autonomous Vehicles}.
\newblock \bibinfo{journal}{\emph{Nature}} \bibinfo{volume}{615}, \bibinfo{number}{7953} (\bibinfo{date}{March} \bibinfo{year}{2023}), \bibinfo{pages}{620--627}.
\newblock
\showISSN{1476-4687}
\urldef\tempurl%
\url{https://doi.org/10.1038/s41586-023-05732-2}
\showDOI{\tempurl}


\bibitem[Gambi et~al\mbox{.}(2019)]%
        {gene-data-1}
\bibfield{author}{\bibinfo{person}{Alessio Gambi}, \bibinfo{person}{Tri Huynh}, {and} \bibinfo{person}{Gordon Fraser}.} \bibinfo{year}{2019}\natexlab{}.
\newblock \showarticletitle{Generating effective test cases for self-driving cars from police reports}. In \bibinfo{booktitle}{\emph{Proceedings of the 2019 27th ACM Joint Meeting on European Software Engineering Conference and Symposium on the Foundations of Software Engineering}} (Tallinn, Estonia) \emph{(\bibinfo{series}{ESEC/FSE 2019})}. \bibinfo{publisher}{Association for Computing Machinery}, \bibinfo{address}{New York, NY, USA}, \bibinfo{pages}{257–267}.
\newblock
\showISBNx{9781450355728}
\urldef\tempurl%
\url{https://doi.org/10.1145/3338906.3338942}
\showDOI{\tempurl}


\bibitem[Geiger and Heckerman(1994)]%
        {bdeu}
\bibfield{author}{\bibinfo{person}{Dan Geiger} {and} \bibinfo{person}{David Heckerman}.} \bibinfo{year}{1994}\natexlab{}.
\newblock \showarticletitle{Learning Gaussian Networks}.
\newblock In \bibinfo{booktitle}{\emph{Uncertainty in Artificial Intelligence}}, \bibfield{editor}{\bibinfo{person}{Ramon~Lopez {de Mantaras}} {and} \bibinfo{person}{David Poole}} (Eds.). \bibinfo{publisher}{Morgan Kaufmann}, \bibinfo{address}{San Francisco (CA)}, \bibinfo{pages}{235--243}.
\newblock
\showISBNx{978-1-55860-332-5}
\urldef\tempurl%
\url{https://doi.org/10.1016/B978-1-55860-332-5.50035-3}
\showDOI{\tempurl}


\bibitem[Guo et~al\mbox{.}(2020)]%
        {safe_to_drive_metric}
\bibfield{author}{\bibinfo{person}{Junyao Guo}, \bibinfo{person}{Unmesh Kurup}, {and} \bibinfo{person}{Mohak Shah}.} \bibinfo{year}{2020}\natexlab{}.
\newblock \showarticletitle{Is it Safe to Drive? An Overview of Factors, Metrics, and Datasets for Driveability Assessment in Autonomous Driving}.
\newblock \bibinfo{journal}{\emph{IEEE Transactions on Intelligent Transportation Systems}} \bibinfo{volume}{21}, \bibinfo{number}{8} (\bibinfo{year}{2020}), \bibinfo{pages}{3135--3151}.
\newblock
\urldef\tempurl%
\url{https://doi.org/10.1109/TITS.2019.2926042}
\showDOI{\tempurl}


\bibitem[Gutierrez-Osorio and Pedraza(2020)]%
        {data_source_GUTIERREZOSORIO2020432}
\bibfield{author}{\bibinfo{person}{Camilo Gutierrez-Osorio} {and} \bibinfo{person}{César Pedraza}.} \bibinfo{year}{2020}\natexlab{}.
\newblock \showarticletitle{Modern data sources and techniques for analysis and forecast of road accidents: A review}.
\newblock \bibinfo{journal}{\emph{Journal of Traffic and Transportation Engineering (English Edition)}} \bibinfo{volume}{7}, \bibinfo{number}{4} (\bibinfo{year}{2020}), \bibinfo{pages}{432--446}.
\newblock
\showISSN{2095-7564}
\urldef\tempurl%
\url{https://doi.org/10.1016/j.jtte.2020.05.002}
\showDOI{\tempurl}


\bibitem[Huai et~al\mbox{.}(2023)]%
        {huaiDoppelgangerTestGeneration2023}
\bibfield{author}{\bibinfo{person}{Yuqi Huai}, \bibinfo{person}{Yuntianyi Chen}, \bibinfo{person}{Sumaya Almanee}, \bibinfo{person}{Tuan Ngo}, \bibinfo{person}{Xiang Liao}, \bibinfo{person}{Ziwen Wan}, \bibinfo{person}{Qi~Alfred Chen}, {and} \bibinfo{person}{Joshua Garcia}.} \bibinfo{year}{2023}\natexlab{}.
\newblock \showarticletitle{Doppelg{\"a}nger {{Test Generation}} for {{Revealing Bugs}} in {{Autonomous Driving Software}}}. In \bibinfo{booktitle}{\emph{2023 {{IEEE}}/{{ACM}} 45th {{International Conference}} on {{Software Engineering}} ({{ICSE}})}}. \bibinfo{pages}{2591--2603}.
\newblock
\showISSN{1558-1225}
\urldef\tempurl%
\url{https://doi.org/10.1109/ICSE48619.2023.00216}
\showDOI{\tempurl}


\bibitem[{ISO 26262}(2018)]%
        {ISO26262}
\bibfield{author}{\bibinfo{person}{{ISO 26262}}.} \bibinfo{year}{2018}\natexlab{}.
\newblock \bibinfo{title}{Road Vehicles Functional Safety}.
\newblock
\newblock
\urldef\tempurl%
\url{https://www.iso.org/standard/68383.html}
\showURL{%
\tempurl}


\bibitem[{ISO 34501}(2022)]%
        {ISO3450}
\bibfield{author}{\bibinfo{person}{{ISO 34501}}.} \bibinfo{year}{2022}\natexlab{}.
\newblock \bibinfo{title}{Test scenarios for automated driving systems Vocabulary}.
\newblock
\newblock
\urldef\tempurl%
\url{https://www.iso.org/standard/78950.html}
\showURL{%
\tempurl}


\bibitem[{ISO/PAS 21448}(2022)]%
        {ISO21448}
\bibfield{author}{\bibinfo{person}{{ISO/PAS 21448}}.} \bibinfo{year}{2022}\natexlab{}.
\newblock \bibinfo{title}{Road Vehicles Safety of the Intended Functionality}.
\newblock
\newblock
\urldef\tempurl%
\url{https://www.iso.org/standard/77490.html}
\showURL{%
\tempurl}


\bibitem[Jokela et~al\mbox{.}(2019)]%
        {sensing_app9112341}
\bibfield{author}{\bibinfo{person}{Maria Jokela}, \bibinfo{person}{Matti Kutila}, {and} \bibinfo{person}{Pasi Pyykönen}.} \bibinfo{year}{2019}\natexlab{}.
\newblock \showarticletitle{Testing and Validation of Automotive Point-Cloud Sensors in Adverse Weather Conditions}.
\newblock \bibinfo{journal}{\emph{Applied Sciences}} \bibinfo{volume}{9}, \bibinfo{number}{11} (\bibinfo{year}{2019}).
\newblock
\showISSN{2076-3417}
\urldef\tempurl%
\url{https://doi.org/10.3390/app9112341}
\showDOI{\tempurl}


\bibitem[Khatun et~al\mbox{.}(2020)]%
        {khatunScenariobasedExtendedHARA2020}
\bibfield{author}{\bibinfo{person}{Marzana Khatun}, \bibinfo{person}{Michael Gla{\ss}}, {and} \bibinfo{person}{Rolf Jung}.} \bibinfo{year}{2020}\natexlab{}.
\newblock \bibinfo{booktitle}{\emph{Scenario-Based {{Extended HARA Incorporating Functional Safety}} \& {{SOTIF}} for {{Autonomous Driving}}}}.
\newblock 59 pages.
\newblock
\urldef\tempurl%
\url{https://doi.org/10.3850/978-981-14-8593-0_5225-cd}
\showDOI{\tempurl}


\bibitem[Kim et~al\mbox{.}(2022)]%
        {kimDriveFuzzDiscoveringAutonomous2022}
\bibfield{author}{\bibinfo{person}{Seulbae Kim}, \bibinfo{person}{Major Liu}, \bibinfo{person}{Junghwan~"John" Rhee}, \bibinfo{person}{Yuseok Jeon}, \bibinfo{person}{Yonghwi Kwon}, {and} \bibinfo{person}{Chung~Hwan Kim}.} \bibinfo{year}{2022}\natexlab{}.
\newblock \showarticletitle{{{DriveFuzz}}: {{Discovering Autonomous Driving Bugs}} through {{Driving Quality-Guided Fuzzing}}}. In \bibinfo{booktitle}{\emph{Proceedings of the 2022 {{ACM SIGSAC Conference}} on {{Computer}} and {{Communications Security}}}} \emph{(\bibinfo{series}{{{CCS}} '22})}. \bibinfo{publisher}{{Association for Computing Machinery}}, \bibinfo{address}{{New York, NY, USA}}, \bibinfo{pages}{1753--1767}.
\newblock
\showISBNx{978-1-4503-9450-5}
\urldef\tempurl%
\url{https://doi.org/10.1145/3548606.3560558}
\showDOI{\tempurl}


\bibitem[Li et~al\mbox{.}(2020)]%
        {liAVFUZZERFindingSafety2020}
\bibfield{author}{\bibinfo{person}{Guanpeng Li}, \bibinfo{person}{Yiran Li}, \bibinfo{person}{Saurabh Jha}, \bibinfo{person}{Timothy Tsai}, \bibinfo{person}{Michael Sullivan}, \bibinfo{person}{Siva Kumar~Sastry Hari}, \bibinfo{person}{Zbigniew Kalbarczyk}, {and} \bibinfo{person}{Ravishankar Iyer}.} \bibinfo{year}{2020}\natexlab{}.
\newblock \showarticletitle{{{AV-FUZZER}}: {{Finding Safety Violations}} in {{Autonomous Driving Systems}}}. In \bibinfo{booktitle}{\emph{2020 {{IEEE}} 31st {{International Symposium}} on {{Software Reliability Engineering}} ({{ISSRE}})}}. \bibinfo{pages}{25--36}.
\newblock
\showISSN{2332-6549}
\urldef\tempurl%
\url{https://doi.org/10.1109/ISSRE5003.2020.00012}
\showDOI{\tempurl}


\bibitem[Li et~al\mbox{.}(2023)]%
        {liDatadrivenBayesianNetwork2023}
\bibfield{author}{\bibinfo{person}{Huanhuan Li}, \bibinfo{person}{Xujie Ren}, {and} \bibinfo{person}{Zaili Yang}.} \bibinfo{year}{2023}\natexlab{}.
\newblock \showarticletitle{Data-Driven {{Bayesian}} Network for Risk Analysis of Global Maritime Accidents}.
\newblock \bibinfo{journal}{\emph{Reliability Engineering \& System Safety}}  \bibinfo{volume}{230} (\bibinfo{date}{Feb.} \bibinfo{year}{2023}), \bibinfo{pages}{108938}.
\newblock
\showISSN{0951-8320}
\urldef\tempurl%
\url{https://doi.org/10.1016/j.ress.2022.108938}
\showDOI{\tempurl}


\bibitem[Ma et~al\mbox{.}(2023)]%
        {maValidationVerificationDomainspecific2023}
\bibfield{author}{\bibinfo{person}{Qin Ma}, \bibinfo{person}{Monika {Kaczmarek-He{\ss}}}, {and} \bibinfo{person}{Sybren {de Kinderen}}.} \bibinfo{year}{2023}\natexlab{}.
\newblock \showarticletitle{Validation and Verification in Domain-Specific Modeling Method Engineering: An Integrated Life-Cycle View}.
\newblock \bibinfo{journal}{\emph{Software and Systems Modeling}} \bibinfo{volume}{22}, \bibinfo{number}{2} (\bibinfo{date}{April} \bibinfo{year}{2023}), \bibinfo{pages}{647--666}.
\newblock
\showISSN{1619-1374}
\urldef\tempurl%
\url{https://doi.org/10.1007/s10270-022-01056-3}
\showDOI{\tempurl}


\bibitem[Menzel et~al\mbox{.}(2019)]%
        {menzelFunctionalLogicalScenarios2019}
\bibfield{author}{\bibinfo{person}{Till Menzel}, \bibinfo{person}{Gerrit Bagschik}, \bibinfo{person}{Leon Isensee}, \bibinfo{person}{Andre Schomburg}, {and} \bibinfo{person}{M. Maurer}.} \bibinfo{year}{2019}\natexlab{}.
\newblock \showarticletitle{From {{Functional}} to {{Logical Scenarios}}: {{Detailing}} a {{Keyword-Based Scenario Description}} for {{Execution}} in a {{Simulation Environment}}}.
\newblock \bibinfo{journal}{\emph{2019 IEEE Intelligent Vehicles Symposium (IV)}} (\bibinfo{year}{2019}).
\newblock
\urldef\tempurl%
\url{https://doi.org/10.1109/IVS.2019.8814099}
\showDOI{\tempurl}


\bibitem[Nali{\'c} et~al\mbox{.}(2020)]%
        {nalicScenarioBasedTesting2020}
\bibfield{author}{\bibinfo{person}{{\DJ}emin Nali{\'c}}, \bibinfo{person}{Tomislav Mihalj}, \bibinfo{person}{Maximilian Baeumler}, \bibinfo{person}{Matthias Lehmann}, \bibinfo{person}{Arno Eichberger}, {and} \bibinfo{person}{Stefan Bernsteiner}.} \bibinfo{year}{2020}\natexlab{}.
\newblock \bibinfo{booktitle}{\emph{Scenario {{Based Testing}} of {{Automated Driving Systems}}: {{A Literature Survey}}}}.
\newblock
\urldef\tempurl%
\url{https://doi.org/10.46720/f2020-acm-096}
\showDOI{\tempurl}


\bibitem[Nogueira et~al\mbox{.}(2022a)]%
        {causal_discover_nogueira_methods_2022}
\bibfield{author}{\bibinfo{person}{Ana~Rita Nogueira}, \bibinfo{person}{Andrea Pugnana}, \bibinfo{person}{Salvatore Ruggieri}, \bibinfo{person}{Dino Pedreschi}, {and} \bibinfo{person}{João Gama}.} \bibinfo{year}{2022}\natexlab{a}.
\newblock \showarticletitle{Methods and tools for causal discovery and causal inference}. In \bibinfo{booktitle}{\emph{{WIREs} {Data} {Mining} and {Knowledge} {Discovery}}}, Vol.~\bibinfo{volume}{12}.
\newblock
\urldef\tempurl%
\url{https://doi.org/10.1002/widm.1449}
\showDOI{\tempurl}


\bibitem[Nogueira et~al\mbox{.}(2022b)]%
        {nogueiraMethodsToolsCausal2022}
\bibfield{author}{\bibinfo{person}{Ana~Rita Nogueira}, \bibinfo{person}{Andrea Pugnana}, \bibinfo{person}{Salvatore Ruggieri}, \bibinfo{person}{Dino Pedreschi}, {and} \bibinfo{person}{Jo{\~a}o Gama}.} \bibinfo{year}{2022}\natexlab{b}.
\newblock \showarticletitle{Methods and Tools for Causal Discovery and Causal Inference}.
\newblock \bibinfo{journal}{\emph{WIREs Data Mining and Knowledge Discovery}} \bibinfo{volume}{12}, \bibinfo{number}{2} (\bibinfo{year}{2022}), \bibinfo{pages}{e1449}.
\newblock
\showISSN{1942-4795}
\urldef\tempurl%
\url{https://doi.org/10.1002/widm.1449}
\showDOI{\tempurl}


\bibitem[Pearl and Mackenzie(2018)]%
        {bookPearlMackenzie18}
\bibfield{author}{\bibinfo{person}{Judea Pearl} {and} \bibinfo{person}{Dana Mackenzie}.} \bibinfo{year}{2018}\natexlab{}.
\newblock \bibinfo{booktitle}{\emph{The Book of Why}}.
\newblock \bibinfo{publisher}{Basic Books}, \bibinfo{address}{New York}.
\newblock
\showISBNx{978-0-465-09760-9}


\bibitem[SAE(2018)]%
        {SAEJ3016}
\bibfield{author}{\bibinfo{person}{SAE}.} \bibinfo{year}{2018}\natexlab{}.
\newblock \bibinfo{title}{Taxonomy and Definitions for Terms Related to Driving Automation Systems for on-Road Motor Vehicles}.
\newblock \bibinfo{howpublished}{Online}.
\newblock
\urldef\tempurl%
\url{https://www.sae.org/standards/content/j3016_201806/}
\showURL{%
\tempurl}


\bibitem[Stocco et~al\mbox{.}(2023)]%
        {defectsADSgap}
\bibfield{author}{\bibinfo{person}{Andrea Stocco}, \bibinfo{person}{Brian Pulfer}, {and} \bibinfo{person}{Paolo Tonella}.} \bibinfo{year}{2023}\natexlab{}.
\newblock \showarticletitle{Mind the Gap! A Study on the Transferability of Virtual Versus Physical-World Testing of Autonomous Driving Systems}.
\newblock \bibinfo{journal}{\emph{IEEE Transactions on Software Engineering}} \bibinfo{volume}{49}, \bibinfo{number}{4} (\bibinfo{year}{2023}), \bibinfo{pages}{1928--1940}.
\newblock
\urldef\tempurl%
\url{https://doi.org/10.1109/TSE.2022.3202311}
\showDOI{\tempurl}


\bibitem[Sun et~al\mbox{.}(2023)]%
        {gene_lawbreaker}
\bibfield{author}{\bibinfo{person}{Yang Sun}, \bibinfo{person}{Christopher~M. Poskitt}, \bibinfo{person}{Jun Sun}, \bibinfo{person}{Yuqi Chen}, {and} \bibinfo{person}{Zijiang Yang}.} \bibinfo{year}{2023}\natexlab{}.
\newblock \showarticletitle{LawBreaker: An Approach for Specifying Traffic Laws and Fuzzing Autonomous Vehicles}. In \bibinfo{booktitle}{\emph{Proceedings of the 37th IEEE/ACM International Conference on Automated Software Engineering}} (Rocheste, USA) \emph{(\bibinfo{series}{ASE '22})}. \bibinfo{publisher}{Association for Computing Machinery}, \bibinfo{address}{New York, NY, USA}, Article \bibinfo{articleno}{62}, \bibinfo{numpages}{12}~pages.
\newblock
\showISBNx{9781450394758}
\urldef\tempurl%
\url{https://doi.org/10.1145/3551349.3556897}
\showDOI{\tempurl}


\bibitem[Tang et~al\mbox{.}(2023)]%
        {tangSurveyAutomatedDriving2023a}
\bibfield{author}{\bibinfo{person}{Shuncheng Tang}, \bibinfo{person}{Zhenya Zhang}, \bibinfo{person}{Yi Zhang}, \bibinfo{person}{Jixiang Zhou}, \bibinfo{person}{Yan Guo}, \bibinfo{person}{Shuang Liu}, \bibinfo{person}{Shengjian Guo}, \bibinfo{person}{Yan-Fu Li}, \bibinfo{person}{Lei Ma}, \bibinfo{person}{Yinxing Xue}, {and} \bibinfo{person}{Yang Liu}.} \bibinfo{year}{2023}\natexlab{}.
\newblock \showarticletitle{A {{Survey}} on {{Automated Driving System Testing}}: {{Landscapes}} and {{Trends}}}.
\newblock \bibinfo{journal}{\emph{ACM Transactions on Software Engineering and Methodology}} \bibinfo{volume}{32}, \bibinfo{number}{5} (\bibinfo{date}{July} \bibinfo{year}{2023}), \bibinfo{pages}{124:1--124:62}.
\newblock
\showISSN{1049-331X}
\urldef\tempurl%
\url{https://doi.org/10.1145/3579642}
\showDOI{\tempurl}


\bibitem[Tian et~al\mbox{.}(2018)]%
        {percp_deeptest}
\bibfield{author}{\bibinfo{person}{Yuchi Tian}, \bibinfo{person}{Kexin Pei}, \bibinfo{person}{Suman Jana}, {and} \bibinfo{person}{Baishakhi Ray}.} \bibinfo{year}{2018}\natexlab{}.
\newblock \showarticletitle{DeepTest: automated testing of deep-neural-network-driven autonomous cars}. In \bibinfo{booktitle}{\emph{Proceedings of the 40th International Conference on Software Engineering}} (Gothenburg, Sweden) \emph{(\bibinfo{series}{ICSE '18})}. \bibinfo{publisher}{Association for Computing Machinery}, \bibinfo{address}{New York, NY, USA}, \bibinfo{pages}{303–314}.
\newblock
\showISBNx{9781450356381}
\urldef\tempurl%
\url{https://doi.org/10.1145/3180155.3180220}
\showDOI{\tempurl}


\bibitem[Wang et~al\mbox{.}(2023)]%
        {wangRiskScenarioGeneration2023}
\bibfield{author}{\bibinfo{person}{Tong Wang}, \bibinfo{person}{Xiaohui Kuang}, \bibinfo{person}{Huan Deng}, \bibinfo{person}{Taotao Gu}, \bibinfo{person}{Wei Kong}, \bibinfo{person}{Jianwen Tian}, {and} \bibinfo{person}{Gang Zhao}.} \bibinfo{year}{2023}\natexlab{}.
\newblock \showarticletitle{Risk {{Scenario Generation}} for {{Autonomous Driving Systems}} Based on {{Scenario Evaluation Model}}}. In \bibinfo{booktitle}{\emph{2023 {{International Joint Conference}} on {{Neural Networks}} ({{IJCNN}})}}. \bibinfo{pages}{1--8}.
\newblock
\showISSN{2161-4407}
\urldef\tempurl%
\url{https://doi.org/10.1109/IJCNN54540.2023.10191164}
\showDOI{\tempurl}


\bibitem[Wang et~al\mbox{.}(2022)]%
        {ddpg}
\bibfield{author}{\bibinfo{person}{Xu Wang}, \bibinfo{person}{Sen Wang}, \bibinfo{person}{Xingxing Liang}, \bibinfo{person}{Dawei Zhao}, \bibinfo{person}{Jincai Huang}, \bibinfo{person}{Xin Xu}, \bibinfo{person}{Bin Dai}, {and} \bibinfo{person}{Qiguang Miao}.} \bibinfo{year}{2022}\natexlab{}.
\newblock \showarticletitle{Deep Reinforcement Learning: A Survey}.
\newblock \bibinfo{journal}{\emph{IEEE Transactions on Neural Networks and Learning Systems}} (\bibinfo{year}{2022}), \bibinfo{pages}{1--15}.
\newblock
\urldef\tempurl%
\url{https://doi.org/10.1109/TNNLS.2022.3207346}
\showDOI{\tempurl}


\bibitem[Waymo(2024)]%
        {waymoOpenDataset}
\bibfield{author}{\bibinfo{person}{Waymo}.} \bibinfo{year}{2024}\natexlab{}.
\newblock \bibinfo{title}{{O}pen {D}ataset – {W}aymo --- waymo.com}.
\newblock \bibinfo{howpublished}{\url{https://waymo.com/open/}}.
\newblock
\newblock
\shownote{[Accessed 11-01-2024]}.


\bibitem[Wu et~al\mbox{.}(2019)]%
        {resnet}
\bibfield{author}{\bibinfo{person}{Zifeng Wu}, \bibinfo{person}{Chunhua Shen}, {and} \bibinfo{person}{Anton {van den Hengel}}.} \bibinfo{year}{2019}\natexlab{}.
\newblock \showarticletitle{Wider or Deeper: Revisiting the ResNet Model for Visual Recognition}.
\newblock \bibinfo{journal}{\emph{Pattern Recognition}}  \bibinfo{volume}{90} (\bibinfo{year}{2019}), \bibinfo{pages}{119--133}.
\newblock
\showISSN{0031-3203}
\urldef\tempurl%
\url{https://doi.org/10.1016/j.patcog.2019.01.006}
\showDOI{\tempurl}


\bibitem[Zanga et~al\mbox{.}(2022)]%
        {causal_discover}
\bibfield{author}{\bibinfo{person}{Alessio Zanga}, \bibinfo{person}{Elif Ozkirimli}, {and} \bibinfo{person}{Fabio Stella}.} \bibinfo{year}{2022}\natexlab{}.
\newblock \showarticletitle{A Survey on Causal Discovery: Theory and Practice}.
\newblock \bibinfo{journal}{\emph{International Journal of Approximate Reasoning}}  \bibinfo{volume}{151} (\bibinfo{year}{2022}), \bibinfo{pages}{101--129}.
\newblock
\showISSN{0888-613X}
\urldef\tempurl%
\url{https://doi.org/10.1016/j.ijar.2022.09.004}
\showDOI{\tempurl}


\bibitem[Zhang et~al\mbox{.}(2018)]%
        {percp_deeproad}
\bibfield{author}{\bibinfo{person}{Mengshi Zhang}, \bibinfo{person}{Yuqun Zhang}, \bibinfo{person}{Lingming Zhang}, \bibinfo{person}{Cong Liu}, {and} \bibinfo{person}{Sarfraz Khurshid}.} \bibinfo{year}{2018}\natexlab{}.
\newblock \showarticletitle{DeepRoad: GAN-based metamorphic testing and input validation framework for autonomous driving systems}. In \bibinfo{booktitle}{\emph{Proceedings of the 33rd ACM/IEEE International Conference on Automated Software Engineering}} (Montpellier, France) \emph{(\bibinfo{series}{ASE '18})}. \bibinfo{publisher}{Association for Computing Machinery}, \bibinfo{address}{New York, NY, USA}, \bibinfo{pages}{132–142}.
\newblock
\showISBNx{9781450359375}
\urldef\tempurl%
\url{https://doi.org/10.1145/3238147.3238187}
\showDOI{\tempurl}


\bibitem[Zhou et~al\mbox{.}(2020)]%
        {percep_deepbillbord}
\bibfield{author}{\bibinfo{person}{Husheng Zhou}, \bibinfo{person}{Wei Li}, \bibinfo{person}{Zelun Kong}, \bibinfo{person}{Junfeng Guo}, \bibinfo{person}{Yuqun Zhang}, \bibinfo{person}{Bei Yu}, \bibinfo{person}{Lingming Zhang}, {and} \bibinfo{person}{Cong Liu}.} \bibinfo{year}{2020}\natexlab{}.
\newblock \showarticletitle{DeepBillboard: systematic physical-world testing of autonomous driving systems}. In \bibinfo{booktitle}{\emph{Proceedings of the ACM/IEEE 42nd International Conference on Software Engineering}} (Seoul, South Korea) \emph{(\bibinfo{series}{ICSE '20})}. \bibinfo{publisher}{Association for Computing Machinery}, \bibinfo{address}{New York, NY, USA}, \bibinfo{pages}{347–358}.
\newblock
\showISBNx{9781450371216}
\urldef\tempurl%
\url{https://doi.org/10.1145/3377811.3380422}
\showDOI{\tempurl}


\end{thebibliography}


\received{Day month 2024}
\received[revised]{Day Month 2024}
\received[accepted]{Day Month 2024}

\end{document}